\documentclass[referee]{raa}           
\usepackage{graphicx,times}
\usepackage{natbib}
\usepackage{amssymb,amsmath}
\usepackage{epstopdf}

\bibpunct{(}{)}{;}{a}{}{,}

\usepackage{caption}
\usepackage[papersize={9in,11in}]{geometry}
\usepackage{hyperref}
\hypersetup{pdftitle = The title of my PDF, pdfauthor = My name, pdfsubject= The subject, pdfkeywords = keyword1 keyword2 keyword3} 
\hypersetup{colorlinks = true, linkcolor = green, anchorcolor = red, citecolor = blue, filecolor = red, pagecolor = red, urlcolor = red}

\begin{document}

   \title{A Study of Magnetized White Dwarf + Helium Star Binary Evolution to Type Ia Supernovae
}

 \volnopage{ {\bf 20XX} Vol.\ {\bf X} No. {\bf XX}, 000--000}
   \setcounter{page}{1}

   \author{Zhe Cui\inst{1}, Xiang-Dong Li\inst{1,2}
   }

   \institute{ School of Astronomy and Space Science, Nanjing University, Nanjing 210023, China;
   	{\it lixd@nju.edu.cn}\\
   	\and
   	Key Laboratory of Modern Astronomy and Astrophysics, Nanjing University,
   	Ministry of Education, Nanjing 210023, China\\
\vs \no
   {\small Received 20XX Month Day; accepted 20XX Month Day}
}

\abstract{
	The white dwarf (WD) + helium (He) star binary channel plays an important role in the single degenerate scenario for the progenitors of type Ia supernovae (SNe Ia). Previous studies on the WD + main sequence star evolution have shown that the magnetic fields of WDs may significantly influence their accretion and nuclear burning processes.
	In this work we focus on the evolution of magnetized WD + He star binaries with detailed stellar evolution and binary population synthesis (BPS) calculations. In the case of magnetized WDs, the magnetic fields may disrupt the inner regions of the accretion disk, funnel the accretion flow onto the polar caps, and even confine helium burning within the caps. We find that, for WDs with sufficiently strong magnetic fields, the parameter space of the potential SN Ia progenitor systems shrinks toward shorter orbital periods and lower donor masses compared with that in the non-magnetized WD case. The reason is that the magnetic confinement usually works with relatively high mass transfer rates, which can trigger strong wind mass loss from the WD, thus limiting the He-rich mass accumulation efficiency. The surviving companion stars are likely of low-mass at the moment of the SN explosions, which can be regarded as a possible explanation for the non-detection of surviving companions after the SNe or inside the SN remnants. However, the corresponding birthrate of Galactic SNe Ia in our high-magnetic models is estimated to be $\sim(0.08-0.13) \times 10^{-3}$yr$^{-1}$ ($\sim 0.17-0.28 \times10^{-3}$yr$^{-1}$ for the non-magnetic models), significantly lower than the observed Galactic SN Ia birthrate.
}

   \authorrunning{Zhe Cui \& X.-D. Li}            
   \titlerunning{A Study of Magnetized White Dwarf + Helium Star Binary Evolution to Type Ia Supernovae}  
   \maketitle

%
\section{Introduction}
\label{sec:intro}

Type Ia supernovae (SNe Ia) are among the most violent events in the universe, which present us the possibilities to probe the evolutionary history of cosmic expansion over the past ten billion years \citep{riess1998,perlmutter1999}. They are regarded as the standard candles for cosmological measurements benefiting from their unified light curves, to derive the values of cosmological parameters (such as the mass density $\Omega_M$, the dark energy density $\Omega_\Lambda$ and the Hubble constant $H_{0}$) \citep{branch1992,hamuy1993,hamuy1996,nomoto1997}, setting off an upsurge in the study of dark energy \citep{riess2007,alam2007,rest2014,demianski2019}.

Spectroscopic and photometric studies prefer SNe Ia to be the runaway thermonuclear fusion of carbon and oxygen in white dwarfs (WDs) once their masses reach the Chandrasekhar mass ($M_{\rm ch}$) limit \citep{Hoeflich1996,nugent1997}. However, the nature of their progenitors is still under debate, and the proposed progenitor scenarios can only explain part of the observations, such as the properties of the host galaxies, the delay times, the birthrates, and the surviving companion stars, etc \citep{cappellaro1997a,iwamoto1999,mannucci2005,mazzali2007,maoz2011,wang2012,ruiz2014,maeda2016,soker2017,patat2018}. Several progenitor scenarios are currently under discussion, but are all confronted with an increasing number of challenges \citep{maoz2012}.

In the so-called double degenerate (DD) scenario, two CO WDs undergo a dynamical coalescence \citep{iben1984} owing to orbital shrinkage caused by gravitational wave radiation, resulting in a combined mass exceeding $M_{\rm ch}$ and a subsequent SN Ia. This scenario is supported by some observational facts such as the absence of H emission lines in the early- and nebular-phase spectra of SNe Ia \citep{leonard2007,brown2012,shappee2013,lundqvist2015,olling2015,sand2018,dimitriadis2019,tucker2019}, and the non-detection of a surviving companion star in the relatively close supernova remnants (SNRs) \citep{schaefer2012,kerzendorf2013,kerzendorf2018,ruiz2018}. Besides, the predicted occurrence rates and delay time distribution (DTD) of SNe Ia from the DD scenario are consistent with observations in both young and old stellar populations \citep{ruiter2009,maoz2010,yungelson2017,liu2016,liu2018}. However, the coalescence of double WDs may also trigger an off-center convective carbon-burning, yielding a high-mass O/Ne WD or collapse to a neutron star, rather than an SN Ia \citep{nomoto1985,shen2012}.

Alternatively, the single degenerate (SD) scenario \citep{whelan1973} assumes that a CO WD accumulates hydrogen- and/or helium-rich material on its surface by rapidly accreting from its non-degenerate companion, which is either a main-sequence (MS) star, a sub-giant star, a redgiant (RG) star, or a He star. Under specific conditions, the accreted material can burn steadily on the surface of the WD, which will eventually explode as a SN Ia when its mass reaches $M_{\rm ch}$ \citep{nomoto1982,hachisu1996,li1997,han2004,lv2009,wang2012}. This scenario is capable of explaining the observational homogeneity of SNe Ia \citep{hamuy1991}, since the SNe Ia realized through this channel are the thermonuclear explosions of similar-mass CO WDs. But the diversities in the maximum luminosities, the light-curve shapes, and the spectra of SNe Ia would resort to other evolutionary channels \citep{phillips1993,branch1996,hamuy1996,umeda1999}. For the SD scenario, the mass transfer through Roche-lobe overflow (RLOF) is crucial since the WD can accumulate mass only when the accretion rate $\dot{M}_{\rm acc}$ is within a narrow range for steady hydrogen and helium burning \citep[e.g.,][]{nomoto1982,kato2004,brooks2016}. The predicted SN Ia birthrates (a few $\times 10^{-4}\,{\rm yr}^{-1})$ \citep{ruiter2009,wang2017} are an order of magnitude lower than the observed ones (a few $\times 10^{-3}\,\rm{{\rm yr}^{-1}}$) \citep{cappellaro1997b,patat2018}. Moreover, the observations of nearby SNe Ia and SNRs provide an upper limit on the luminosity of the surviving companion star, setting stringent constraints on the parameter space of the progenitor binaries \citep{badenes2007,maoz2008,Kerzendorf2009,li2011,kelly2014,maoz2014,ruiz2014}.

Magnetism occupies a non-negligible incidence in observed WDs. With the advent of Sloan Digital Sky Survey and other large-scale spectroscopic surveys, the numbers of known single and binary magnetic WDs (MWDs) with magnetic fields $B$ in the range $\sim10^6- 10^9$ G have increased to more than 600 and 200, respectively  \citep{wickramasinghe2000,york2000,gansicke2002,Schmidt2003,Vanlandingham2005,kulebi2009,sion2014,kepler2015,ferrario2015,ferrario2020}.
Magnetic cataclysmic variables (MCVs) make up about $25\%$ of the known CVs in the magnitude-limited samples and even as high as $36\%$ within 150 pc \citep{ferrario2015,pala2020}.
Observations and studies of MCVs and super-soft X-ray sources (SSSs) have revealed the influence of magnetic field on the WD binary evolution \citep{osborne2001,ferrario2015}. For instance, MCVs generally behave as much stronger X-ray emitters than non-mangnetic ones.
The magnetic fields of WDs have also been taken into account in the study of the SN Ia formation mechanisms. \cite{neunteufel2017} indicates that helium ignition in (rotating) weakly magnetized CO WD may lead to fast and faint hydrogen-free SN explosions. Moreover, since the accreted material is funneled onto a small portion of the WD surface by the magnetic field lines, thermonuclear burning on magnetic WDs could be different from that on non-magnetic WDs even at the same accretion rate. For example, a relatively low accretion rate of $\sim 10^{-9} \,M_{\odot}\,{\rm yr}^{-1}$ may be high enough to sustain local stable hydrogen burning on MWDs \citep{schaefer2010}, implying that magnetism may significantly influence the evolution of accreting WDs (also see discussion in \cite{wheeler2012} and \cite{ablimit2014}). \citet{ablimit2019a,ablimit2019b} investigated the MWD + MS star binary evolution to SNe Ia and showed that, with the magnetic field confinement, the initial parameter spaces of the SN Ia progenitor systems become larger and the surviving companion stars could be dimmer and lighter compared with those in the non-magnetic models, and these features are compatible with non-detection of surviving companion stars in nearby SNe and SNRs as mentioned above. Accretion of He-rich matter on to a WD from its H-depleted donor is related to some interesting events. For example, the He nova V445 Puppis and the X-ray pulsating companion of HD 49798 are both suggested to be the candidates for SN Ia progenitors \citep{kato2008,wang2010b}. In addition, hyper-velocity stars such as US 708 (HVS2) \citep{hirsch2005} may be the surviving donors of SNe Ia from the WD + He star channel \citep{wanghan2009}. As the WD + He star channel also plays an important role in the SD scenario \citep{wang2010a}, it is interesting to examine the influence of magnetism on the evolution of He-accreting MWDs, and this is the objective of our work.

The rest of this paper is organized as follows. We present our methods of simulating the evolution of WD + He binary systems in Section 2, together with some representative examples of MWD + He binary evolution under different magnetic field strengths of the WDs. Section 3 demonstrates the calculated results with an binary population synthesis (BPS) method on the distribution of the progenitor systems and the surviving companions. We summarize our results and conclude in Section 4.

\section{The evolution of WD + He binary systems}
\subsection{Non-magnetic WDs} \label{subsec:nonmag}

In the case of non-magnetic WDs, we simulate the WD + He binary evolution under the optically thick wind model \citep{hachisu1996}, similar as \cite{wang2009}. The mass growth rate $\dot{M}_{\rm wd}$ of the accreting WDs can be expressed as $\dot{M}_{\rm wd} = \eta_{\rm He}\dot{M}_{\rm acc}=\eta_{\rm He}|\dot{M}|$, where $\eta_{\rm He}$ and $\dot{M}$ are the He-rich mass accumulation efficiency and the mass transfer rate, respectively. We use the results of \cite{kato2004} for the $\eta_{\rm He}$ during He-shell flashes, summarized as follows.

(1) If $|\dot{M}|$ exceeds the critical value $\dot{M}_{\rm crit}= 7.2 \times 10^{-6} (M_{\rm wd}/\rm M_\odot -0.6) \rm M_\odot {\rm yr}^{-1}$ \citep{nomoto1982}, $\dot{M}_{\rm wd}$ is limited to $\dot{M}_{\rm crit}$, and the rest of the transferred matter is blown away from the system in the form of isotropic winds at a rate of $\dot{M}_{\rm wind} = |\dot{M}| - \dot{M}_{\rm crit}$, taking the specific angular momentum of the WD. Then $\eta_{\rm He} = \dot{M}_{\rm crit} / |\dot{M}|$.

(2) If $|\dot{M}|$ drops below the minimum accretion rate of weak helium flashes $\dot{M}_{\rm low}$ \citep{woosley1986}, the thermal timescale of the cool and dense WD shell is longer than the accretion-heating timescale, thus the helium shell flashes will be too strong to leave any combustion products. In this situation $\eta_{\rm He} = 0$.

(3) If  $\dot{M}_{\rm stable}<|\dot{M}|<\dot{M}_{\rm crit}$, the accreted helium can burn steadily at a rate of $|\dot{M}|$, thus $\eta_{\rm He} = 1$; if $\dot{M}_{\rm low}<|\dot{M}|<\dot{M}_{\rm stable}$, weak helium flashes will take place with $0 < \eta_{\rm He} < 1$, which is $|\dot{M}|$ and $M_{\rm wd}$ dependent. In this case we use the fitted formulae for $\eta_{\rm He}$ in \citet{kato2004}. Since the maximum accretion rate calculated in \cite{kato2004} is $10^{-5.8} \rm M_\odot {\rm yr}^{-1}$, we simply assume $\eta_{\rm He} = 1$ when $10^{-5.8} {\rm M}_\odot {\rm yr}^{-1}<|\dot{M}|<\dot{M}_{\rm crit}$.

Several groups have performed numerical simulations of He shell burning processes on WDs, and we note that there is not consensus on the He-rich mass accumulation efficiency in the literature. Taking into account the thermal response of He-accreting WDs, \cite{piersanti2013,piersanti2014} made the long-term evolutionary calculations for WDs with initial masses of $0.6-1.1\,\rm M_{\odot}$, and obtained $\eta_{\rm He}$ during different accretion regimes. By introducing the supper-Eddington wind loss, \cite{wang2015} calculated  the values of $\eta_{\rm He}$  with the WD masses ranging from 0.6 to 1.35$\,\rm M_\odot$. The critical accretion rate for stable He-shell burning in \cite{wang2015} is substantially lower than that in \cite{piersanti2013,piersanti2014}. For example, for a 1.2$\,\rm M_\odot$ WD accreting at a rate of $\sim 3.9\times10^{-6}\rm M_{\odot} yr^{-1}$, the He-shell can burn steadily according to \cite{piersanti2013,piersanti2014}, while according to \cite{wang2015} there will be strong supper-Eddington winds from the WD. In their evolutionary calculations, \cite{brooks2016} resolved the full stellar structures of both binary components, and employed the optically thick wind scenario in the wake of the WD's inflating instead of using a form of $\dot{M}_{\rm crit}$ as prescribed by \cite{nomoto1982}. Their calculated values of $\eta_{\rm He}$ are generally between those of \cite{kato2004} and \cite{wang2015}. Assuming that there are supper-Eddington winds during multi-cycle He-shell flashes, \cite{wu2017} also calculated the values of $\eta_{\rm He}$, and found them to be lower than those in \cite{kato2004}.

Considering the uncertainties in $\eta_{\rm He}$ and in order to explore how $\eta_{\rm He}$ influences the evolution of the WD + He binaries, we also make evolutionary calculations by adopting the prescription of \cite{wang2015}. In \cite{wang2015} the optically thick wind in \cite{kato2004} is replaced by the super-Eddington wind, which can significantly reduce $\eta_{\rm He}$. So they can be regarded to represent the high and low ends of the actual values of $\eta_{\rm He}$. In the following we call them the WL and KH prescriptions, respectively.

\subsection{Magnetic WDs} \label{subsec:mag}
For non-magnetic WDs, we assume that the accretion disk extends to the surface of WDs and then the accreted matter is spherically symmetrically distributed on the WDs; but for MWDs with magnetic fields of $B \sim 10^6$ G (for intermediate polars) or $\geq 10^7$ G (for polars), the strong magnetic fields can disrupt part or all of the accretion disk and force the material to fall onto the poles of the WDs. This can be described as polar cap accretion \citep{hameury1986}.

For accreting MWDs with dipolar magnetic fields, the magnetic pressure is expressed as \citep{frank2002}:
\begin{equation}\label{eq1}
	P_{\rm mag} = \frac{(B R_{\rm wd}^3)^2}{8\pi r^6},
\end{equation}
where $r$ is the radial distance from the center of the MWD, and $R_{\rm wd}$ is the radius of the MWD. Equation (1) indicates that $P_{\rm mag}$ increases rapidly as the accreted material approaches the WD's surface. At the magnetospheric radius $R_{\rm M}$, $P_{\rm mag}$ exceeds the ram pressure of the accreting matter, and the motion of the accreted matter starts to be controlled by the magnetic fields. Hence the magnetic confinement condition can be expressed as:
\begin{equation}\label{eq2}
	\frac{(B R_{\rm wd}^3)^2}{8\pi R_{\rm M}^6} \geq \frac{(2GM_{\rm wd})^{1/2}\dot{M}}{4\pi R_{\rm M}^{5/2}}
\end{equation}
or rewritten as:
\begin{equation}\label{eq3}
	B \geq B_{\rm cap} \simeq 9.8 \times 10^3 (\frac{\dot{M}}{10^{-10} \rm M_\odot {\rm yr}^{-1}})^{1/2}(\frac{M_{\rm wd}}{\rm M_\odot})^{1/4}(\frac{R_{\rm wd}}{5 \times 10^8 {\rm cm}})^{-5/4}\, {\rm G},
\end{equation}
where the term on the right-hand-side of Eq.~(\ref{eq2}) is the ram pressure, $G$ is the gravitational constant, and the value of $R_{\rm wd}$ can be calculated from the mass-radius relation of WDs \citep{livio1983}.

To confine the accreted material within the polar caps without spreading over the MWD surface, the magnetic pressure should be larger than the pressure $P_{\rm b}$ at the base of the accreted matter, under which the accreted material can be ignited, so the magnitude of $B$ should satisfy the following requirement:
\begin{equation}\label{eq4}
	B \geq B_{\rm conf} \simeq 9.3 \times 10^7(\frac{R_{\rm wd}}{5 \times 10^8\, {\rm cm}})(\frac{P_{\rm b}}{5 \times 10^{19}\, {\rm dyn cm}^{-2}})^{7/10}(\frac{M_{\rm wd}}{M_\odot})^{-1/2}(\frac{\dot{M}}{10^{-10} M_\odot {\rm yr}^{-1}})^{-1/2}\, {\rm G}.
\end{equation}
For He-shell burning $P_{\rm b} = \frac{GM_{\rm wd}M_{\rm ign}}{4\pi R_{\rm wd}^4}$,  where $M_{\rm ign}$ is the minimum He-shell mass for hydrodynamic He-shell burning.  We calculate the values of $P_{\rm b}$ as a function of $M_{\rm wd}$ from the numerical simulations of He-shell burning by \cite{shen2009}.

In the polar coordinates ($r, \theta$) with the origin located at the WD's center, the dipolar magnetic field lines follow the geometry $r=C\sin^2\theta$, where $C$ is a constant labeling all the field lines emanating from a particular latitude on the WD. The area of the polar cap is estimated to be $\simeq \pi R_{\rm wd}^2 \sin ^2 \beta$, where $\beta$ is the half-angle of the polar cap. The fraction $f_{\rm pc}$ of the area of the two polar caps in the total WD surface area can be expressed to be \citep{frank2002}:
\begin{equation}
	f_{\rm pc} 	\simeq \frac{2 \cdot \pi R_{\rm wd}^2 \sin ^2 \beta}{4\pi R_{\rm wd}^2}
	\simeq \frac{R_{\rm wd} \cos^2 \gamma}{2 R_{\rm M}},
\end{equation}
where $\gamma$ is the angle between the magnetic and rotation axes. In this work, we assume that the WD is an aligned rotator for simplicity, that is, $\gamma=0\degr$. Therefore, the accumulation efficiency $\eta_{\rm He}(\dot{M}, M_{\rm wd})$ is replaced by $\eta_{\rm He}(\dot{M}/f_{\rm pc}, M_{\rm wd})$ in the case of polar cap accretion \citep{ablimit2019a}. We incorporated the above prescriptions into MESA (version r10398) \citep{paxton2015} to follow the evolution of the WD + He star binaries.

\subsection{Binary evolution}

In order to demonstrate the influence of magnetic fields on the WD + He binary evolution, we construct several models with different surface magnetic field strengths $B=$ 0, $3.0\times 10^6 $ and $3.0 \times 10^7$ G. There are observational evidences shown that the surface magnetic fields of the accretors (especially for neutron stars) decayed because of accretion \citep{taam1986}, caused by ohmic dissipation or screen of accreted matter \citep{Geppert1994,Romani1990}. However, whether accretion can induce magnetic field decay in WDs is still debated, considering the fact that polars are strongly magnetized but old WDs that have accreted for a very long time. For instance, AR UMa was classified as a polar in binary system, while the magnetic field of the accreting WD is still as high as $\sim$230 MG \citep{ferrario2002}. Thus, we don't take the magnetic field decay into consideration in our work due to its great uncertainties. In each model, the initial WD masses $M_{\rm wd}^{\rm i}$ are taken to be 0.858, 0.9, 0.95, 1.0, 1.05, 1.1, 1.15, 1.2 and 1.25 $\rm M_\odot$ (the minimum initial mass for WDs that can evolve to SNe Ia is $M_{\rm min}^{\rm i}$ = $0.858 \rm M_\odot$ with the KH prescription, and $M_{\rm min}^{\rm i}$ = $0.9 \rm M_\odot$ with the WL prescription in the case of $B=0$ G), the initial He star masses $M_{2}^{\rm i}$ range from $0.6\,\rm M_\odot$ to $3.5\,\rm M_\odot$ by steps of 0.05 $\rm M_\odot$, and the initial orbital periods $ P^{\rm i} $ (in units of days) vary logarithmically from $ -1.7$ to $2.8$ by steps of 0.1.

In the following we display some representative examples with our binary evolution calculations.

\begin{figure}[htp]
	\centering
	\includegraphics[width=0.495\textwidth]{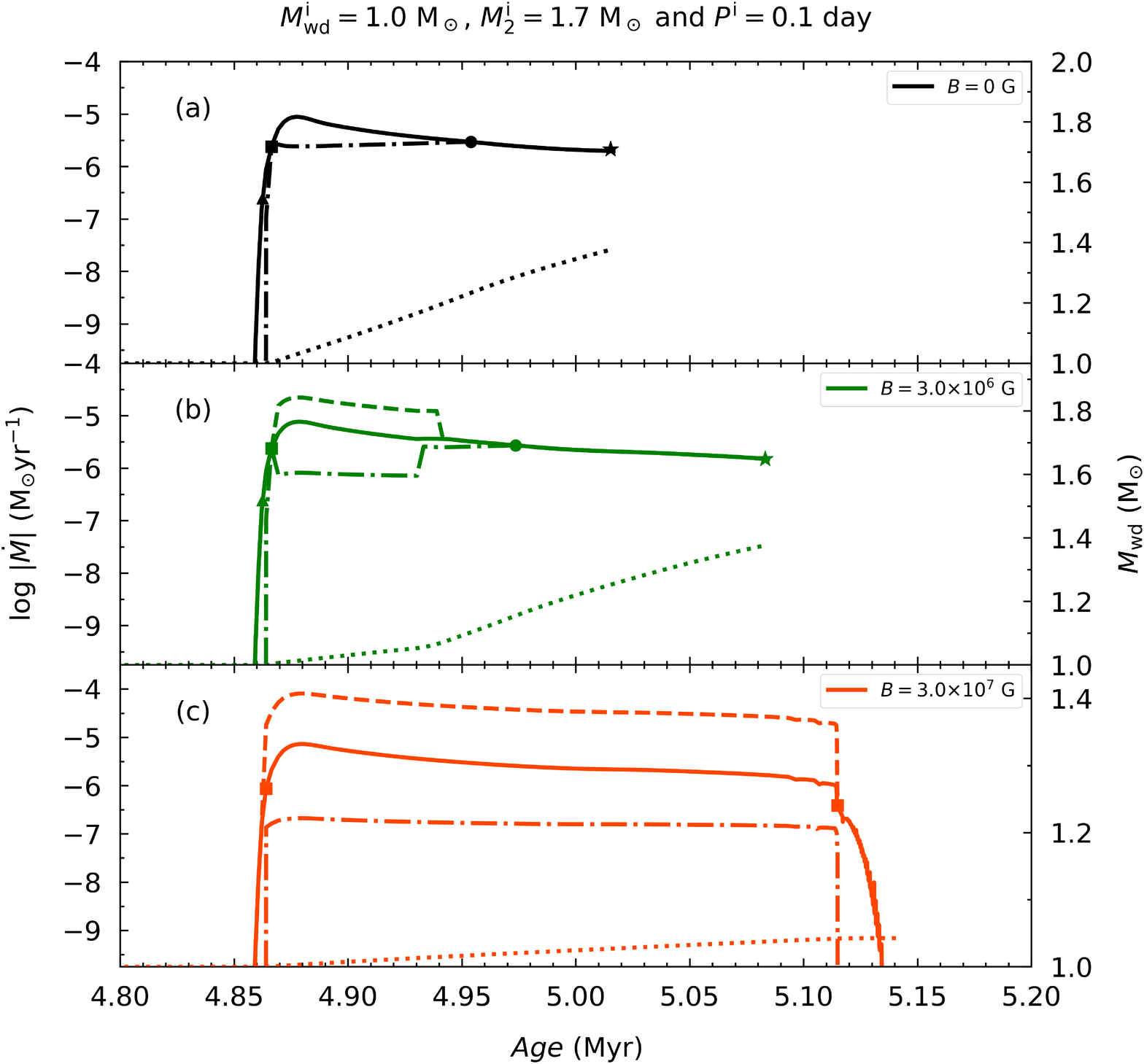}
	\includegraphics[width=0.495\textwidth]{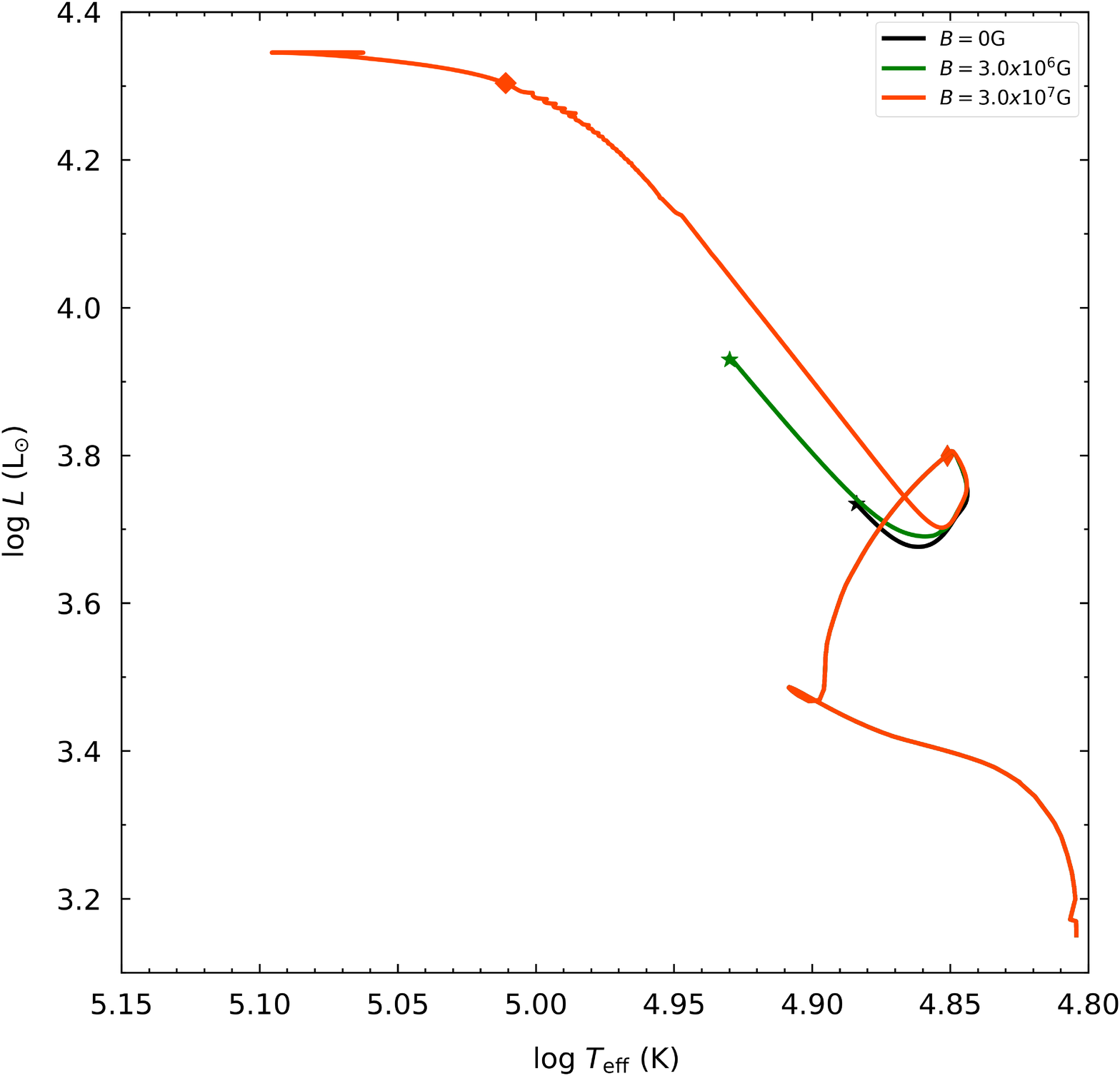}	
	\caption{A representative case of binary evolution with $M_{\rm wd}^{\rm i} = 1.0$ $\rm M_\odot$, $M_{2}^{\rm i} = 1.7$ $\rm M_\odot$ and $P^{\rm i} = 0.1$ day. The WL prescription is adopted for the He-rich mass accumulation efficiency. The left column shows the evolutionary history of the mass transfer rate $\dot{M}$ (the solid lines), WD's mass accumulation rate $\dot{M}_{\rm wd}$ (the dash-dotted lines), $\dot{M}/f_{\rm pc}$ (dashed lines in panel [b]-[c]) and the WD's mass $M_{\rm wd}$ (dotted lines) with different $B$-field strengths. The triangles, circles, and squares mark $|\dot{M}/f_{\rm pc}|=\dot{M}_{\rm low}$, $\dot{M}_{\rm stable}$ and $\dot{M}_{\rm crit}$, respectively. The right panel shows the corresponding evolutionary tracks of the companion star in the HR diagram, with the thin diamond, thick diamond and asterisk symbols denoting the moment of contacting, detaching and SN explosions of systems. \label{fig:f1}}
\end{figure}
Fig.~\ref{fig:f1} shows the results of binary evolution with the initial parameters ($M_{\rm wd}^{\rm i}$/$ \rm M_\odot $, $ M_{2}^{\rm i}/\rm M_\odot $, $ P^{\rm i} $/day) = (1.0, 1.7, 0.1). Panels [a]-[c] in the left column show the results of the non-magnetic ($B=0$ G), intermediate-magnetic ($B=3.0\times 10^6$ G) and high-magnetic ($B=3.0\times 10^7$ G) models, respectively. In the right column, we show the evolution of the companion stars in the H-R diagram, and the moment when the WD is expected to explode in an SN Ia is labeled by an asterisk.

The mass transfer starts at the age of $\sim 4.86$ Myr when the He donor expands and overflows its RL after the exhaustion of central helium. In the non-magnetic model (panel [a]), He-shell burning on the WD is initially unstable because $\dot{M}$ is lower than the weak He-shell flash limit $\dot{M}_{\rm low}$, so violent flashes prevent the mass growth of the WD. With the expansion of the He star's envelope and the contraction of the orbit, $\dot{M}$ increases, the flashes weaken, and eventually the burning becomes thermally stable when $\dot{M} \ge \dot{M}_{\rm stable}$. With further increase in $\dot{M}$, a radiation-driven wind is triggered at the age of $4.865$ Myr, blowing away part of the accreted material. Steady nuclear burning  is accordingly limited at a rate of $\dot{M}_{\rm crit} \simeq 3.2 \times 10^{-6} \rm M_{\odot}yr^{-1}$. The wind phase ceases at the age of $\sim 4.95$ Myr when $\dot{M}$ drops below $\dot{M}_{\rm crit}$. During this steady burning phase the binary can be identified as a SSS until the WD explodes in an SN. In the intermediate-magnetic model (panel [b]), the evolution process is the same as in the non-magnetic one until $\sim 4.865$ Myr, when the magnetic confinement effect commences. Since the transformed rate $\dot{M}/f_{\rm pc}$  is higher than $\dot{M}_{\rm crit}$, the mass accumulation is limited in the pole caps at a rate of $f_{\rm pc}\dot{M}_{\rm crit}$. After about 0.075 Myr, the magnetic confinement fails, and the WD evolves in a similar way as a non-magnetic WD. The WD explodes in the stable He-shell burning phase, 0.07 Myrs later than in the non-magnetic model because of the lower mass growth rate of the WD. The binary parameters at the moment of the SN explosion are ($M_{\rm WD}^{\rm SN}$/$ \rm M_\odot $, $ M_{2}^{\rm SN}/\rm M_\odot $, $ P^{\rm SN} $/day) = (1.378, 1.108, 0.09) in the non-magnetic model, and (1.378, 0.983, 0.098) in the intermediate-magnetic model. The surviving companion star in the intermediate-magnetic model is generally less massive and hotter than in the non-magnetic model, as shown in the right column.

Evolution in the high-magnetic model (panel [c]) is quite different. As the magnetic confinement acts soon after the He-shell ignition until the end of the mass transfer, a substantial fraction of the He-rich matter is blown away from the WD by the supper-Eddington wind, which stops at the age of $\sim 5.12$ Myr. Then $\dot{M}$ rapidly drops below $\dot{M}_{\rm low}$, and strong He-shell flashes expels all the transferred matter. Because of the extensive mass loss, the WD can only grow up to 1.05 $\rm M_\odot $.

\begin{figure}[htp]
	\centering
	\includegraphics[width=0.495\textwidth]{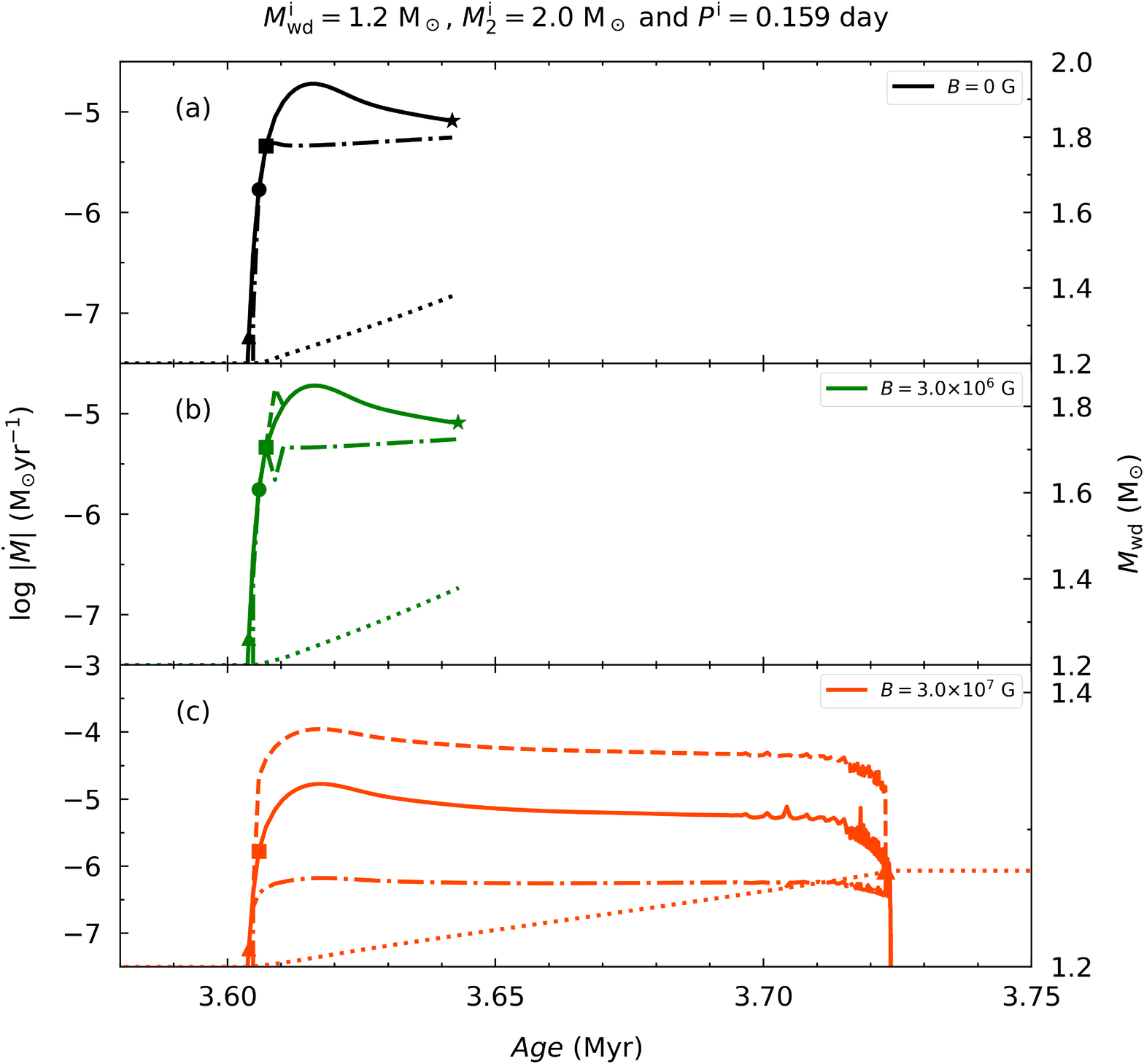}
	\includegraphics[width=0.495\textwidth]{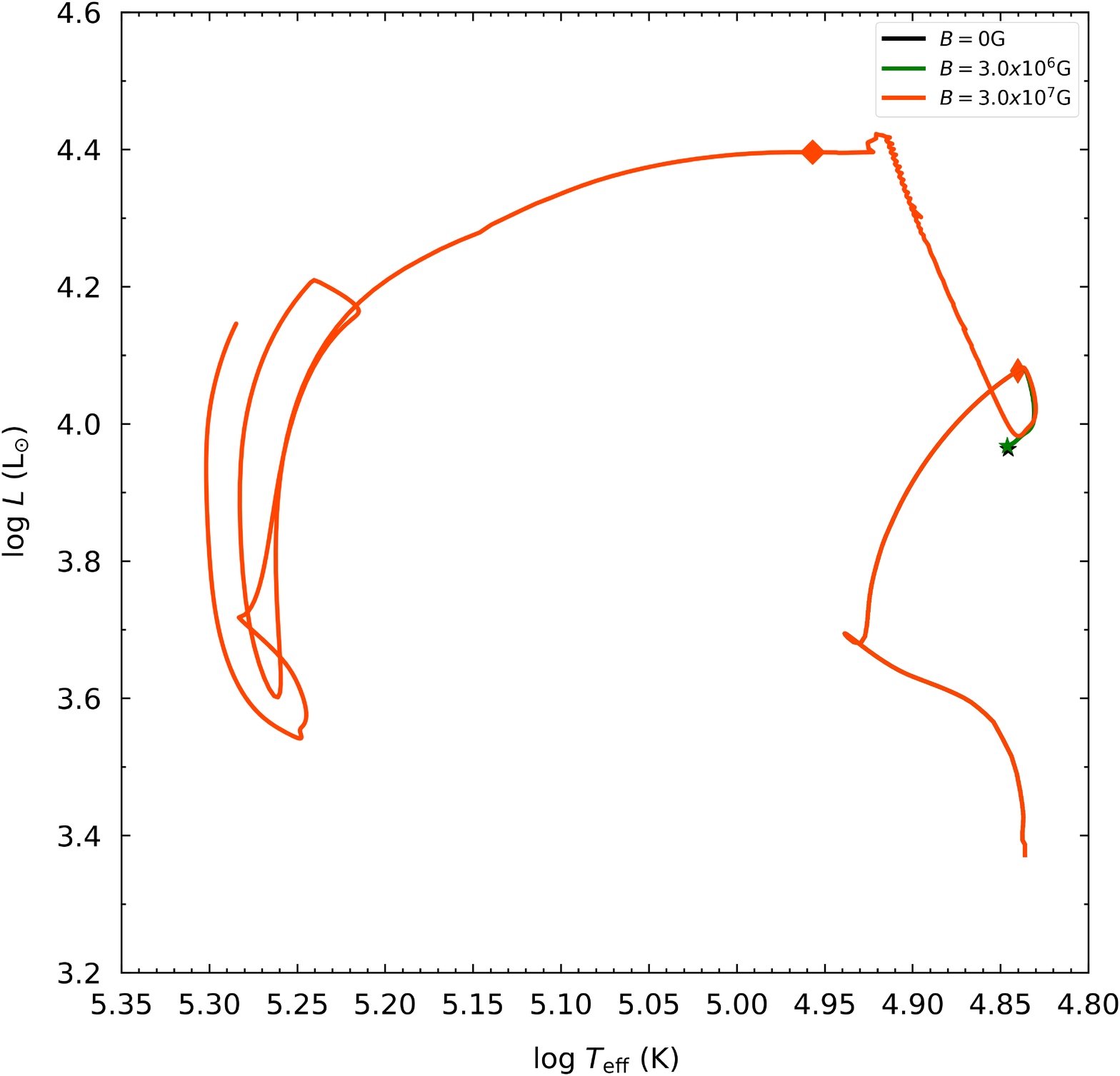}	
	\caption{Same as Fig.~\ref {fig:f1} but for systems with $(M_{\rm wd}^{\rm i}$, $M_{2}^{\rm i}$, $P^{\rm i})=$ (1.2, 2.0, 0.159), and the KH prescription is used for the He-rich mass accumulation efficiency.\label{fig:f2}}
\end{figure}

Fig.~\ref{fig:f2} shows the binary evolution with the initial parameters ($M_{\rm wd}^{\rm i}$/$ \rm M_\odot $, $ M_{2}^{\rm i}/\rm M_\odot $, $ P^{\rm i} $/day) = (1.2, 2.0, 0.159) in the non-magnetic [a] and magnetic ([b] and [c]) models with the KH prescription. Because the donor mass is more massive, the mass transfer rate is higher, and it is more difficult to confine the accreted matter. It can be seen that the evolution in the intermediate-magnetic model is almost the same as in the non-magnetic one, with the magnetic confinement works only for a very short time. The binary parameters at the moment of the SN explosion are ($M_{\rm wd}^{\rm SN}$/$ \rm M_\odot $, $ M_{2}^{\rm SN}/\rm M_\odot $, $ P^{\rm SN} $/day) = (1.378, 1.556, 0.143) and (1.379, 1.546, 0.143) in the non- and intermediate-magnetic models, respectively. In the high-magnetic model, the magnetic confinement always works and $\dot{M}/f_{\rm pc}$ exceeds $\dot{M}_{\rm crit}$ soon after the onset of RLOF and until the mass transfer terminates, so the WD accumulates its mass at a rate of $f_{\rm pc}\dot{M}_{\rm crit}$. Because of the strong wind loss, the WD's mass fails to grow up to $1.378 \rm \rm M_\odot$.

\begin{figure}[htp]
	\centering
	\includegraphics[width=0.48\textwidth]{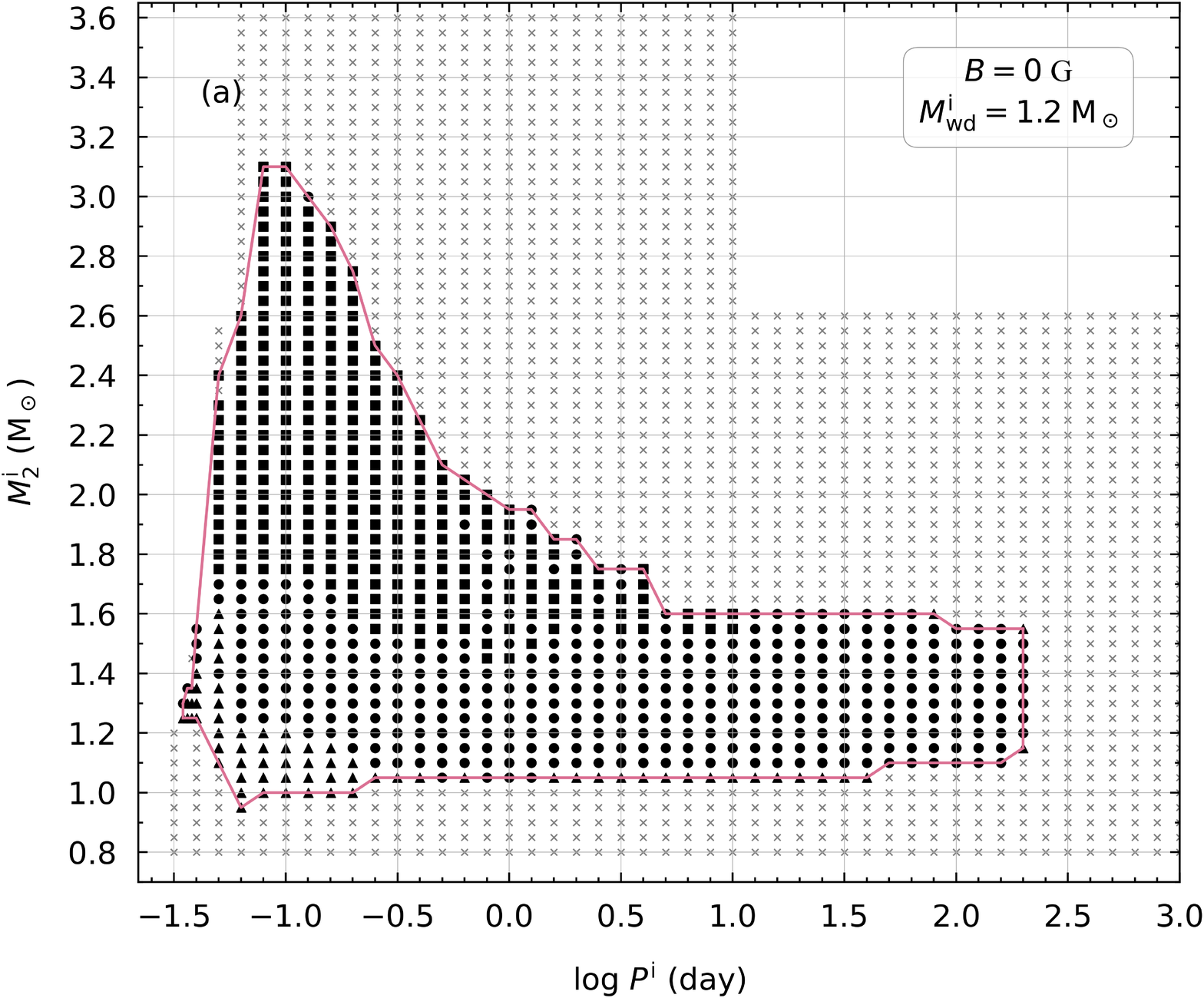}
	\includegraphics[width=0.48\textwidth]{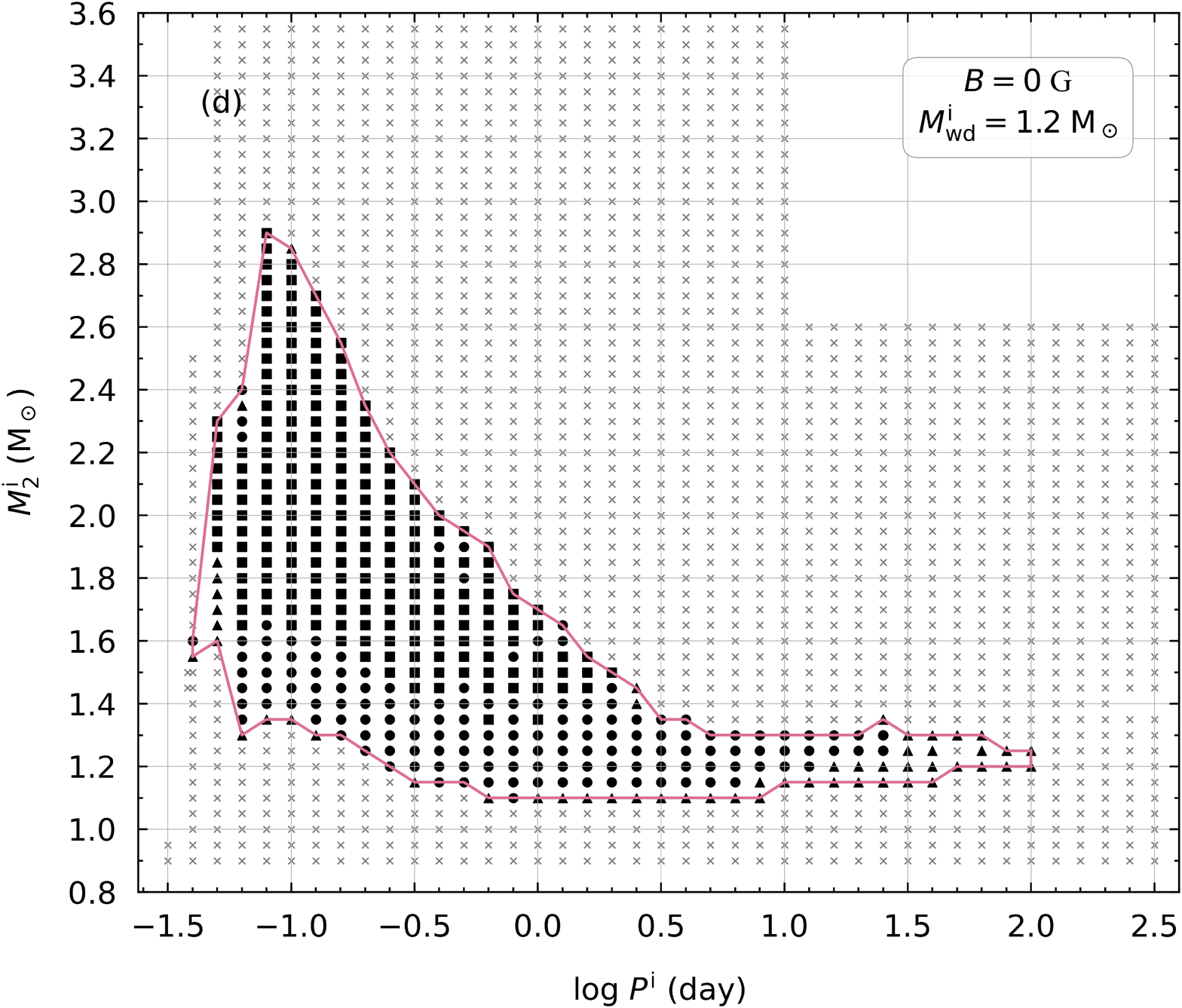}
	\includegraphics[width=0.48\textwidth]{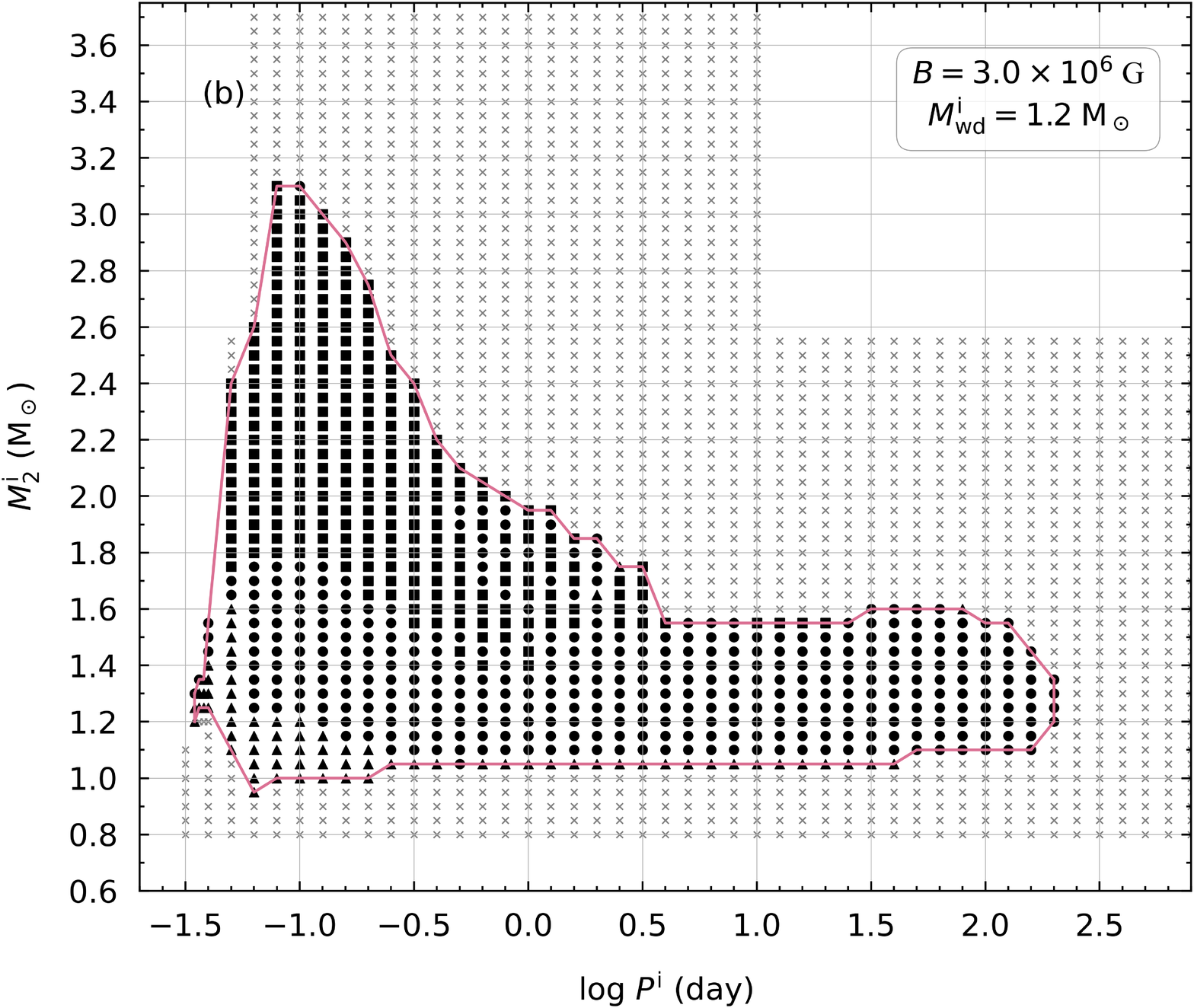}
	\includegraphics[width=0.48\textwidth]{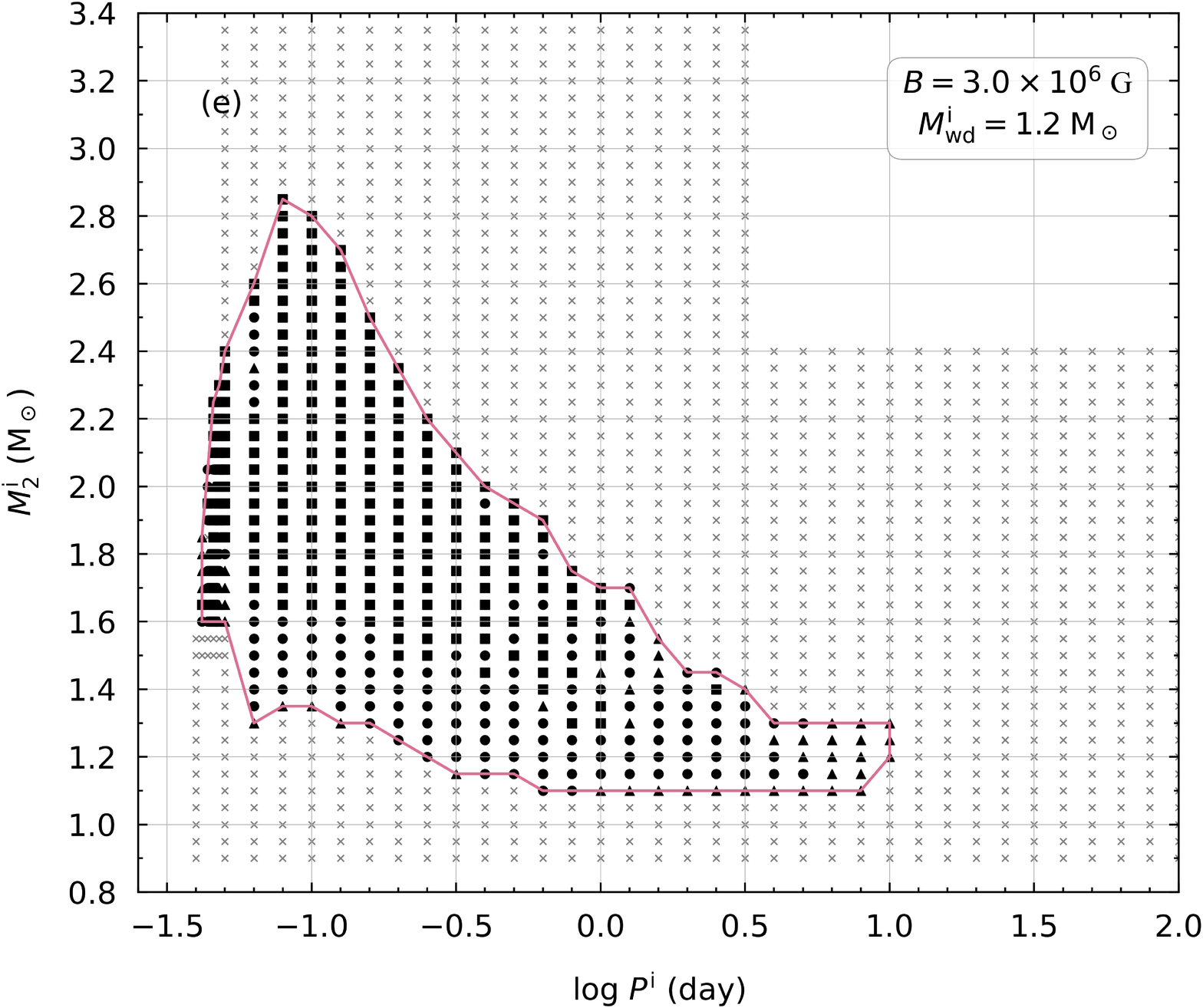}
	\includegraphics[width=0.48\textwidth]{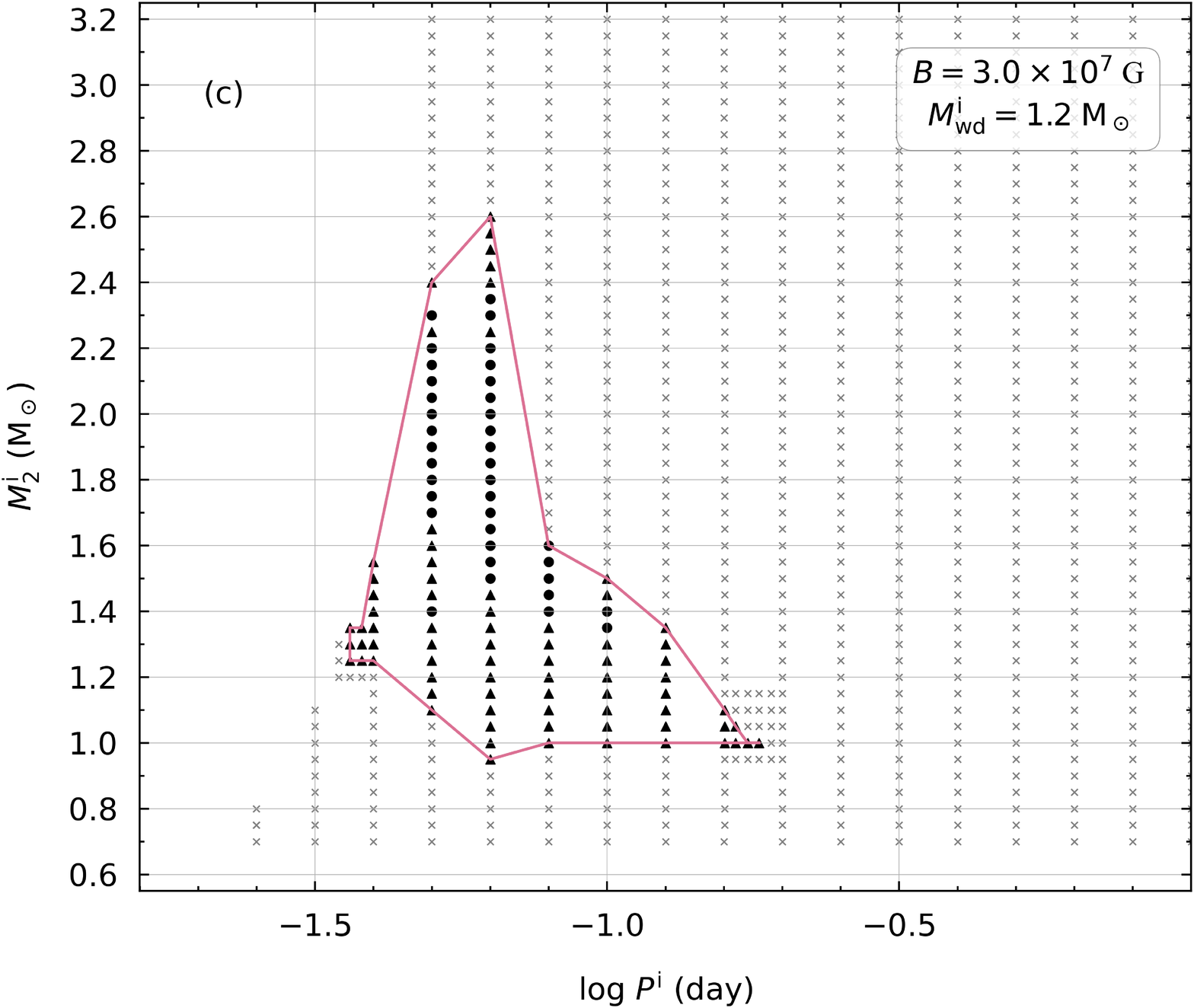}
	\includegraphics[width=0.48\textwidth]{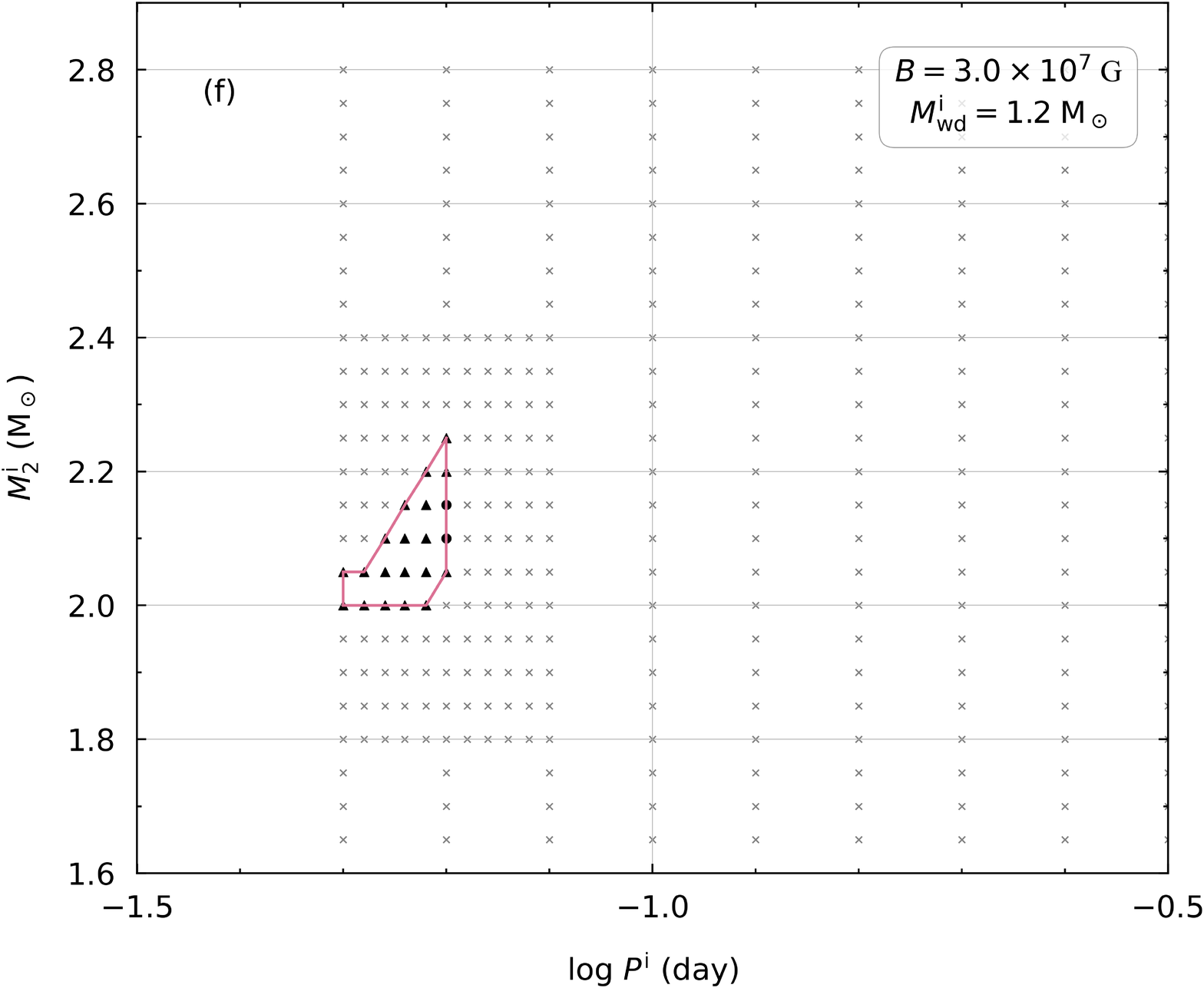}
	\caption{Distributions of the initial orbital period $P^{\rm i}$ and initial donor star mass $M_{\rm 2}^{\rm i}$ for the SN progenitor systems in the WD + He star channel with $M_{\rm wd}^{\rm i} = 1.2$ $\rm M_\odot$. The KH prescription is adopted for the He-rich mass accumulation efficiency in panels [a]-[c]) and the WL prescription in panels [d]-[f]). The three panels in each column correspond to $B = 0$, $3.0 \times 10^6$ and $3.0 \times 10^7$ G, respectively. The triangles, circles and squares denote that the WDs in these systems are experiencing weak He-shell flashes, steady He-shell burning and optically thick wind (KH prescription) or supper-Eddington wind (WL prescription) at the moment of the SN explosion, respectively. The crosses indicate systems that fail to explode owing to strong nova outbursts or wind mass loss from the WDs.\label{fig:f3}}
\end{figure}

Fig.~\ref{fig:f3} summarizes the results of the WD + He star evolution for $M_{\rm wd}^{\rm i} = 1.2 \rm M_\odot$ with different magnetic field strengths and $\eta_{\rm He}$ prescriptions (KH: panels [a-c]; WL: panels [d-f]). Systems with the initial parameters located within the red curves are regarded as potential SN Ia progenitors. The filled triangles, circles and squares denote the WD binaries experiencing weak helium-shell flashes, steady helium-shell burning and optically thick wind at the moment of the SN explosion, respectively, and the crosses indicate systems in which the WDs fail to explode.
We find that the results in the non- and intermediate-magnetic models are almost the same because the $B$-fields in the latter are generally lower than the minimum value of the magnetic field that can confine the matter. In the high-magnetic model, the magnetic confinement takes effects, and the parameter space for the SN progenitors shrinks toward shorter $P^{\rm i}$ and smaller $M_2^{\rm i}$. The reason is that the magnetic confinement is more likely to trigger radiation-driven wind in systems with higher donor masses or longer periods, thus hampering efficient mass accumulation of the WDs in these binaries.

\begin{figure}[htp]
	\centering
	\includegraphics[width=0.48\linewidth]{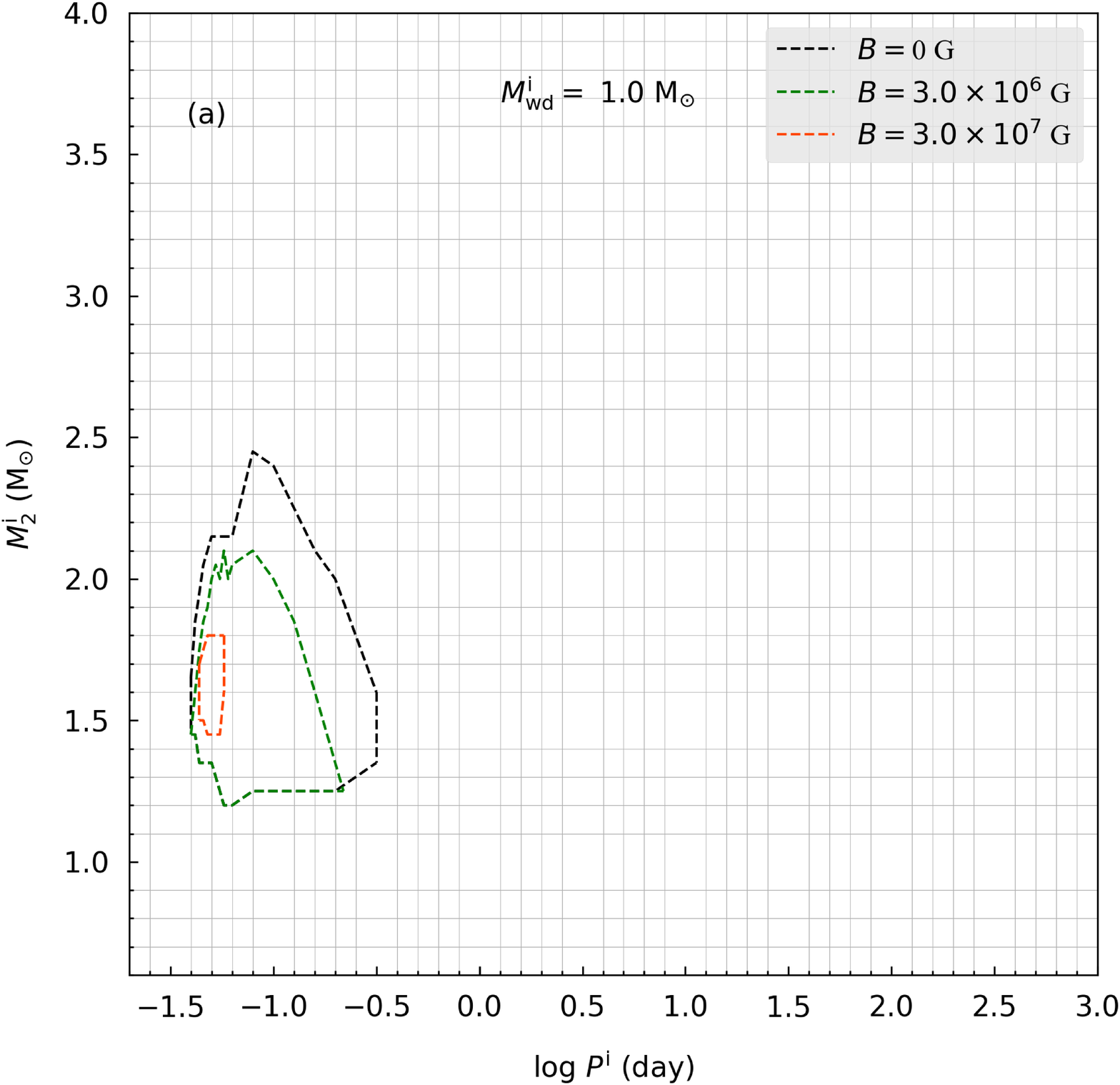}
	\includegraphics[width=0.48\linewidth]{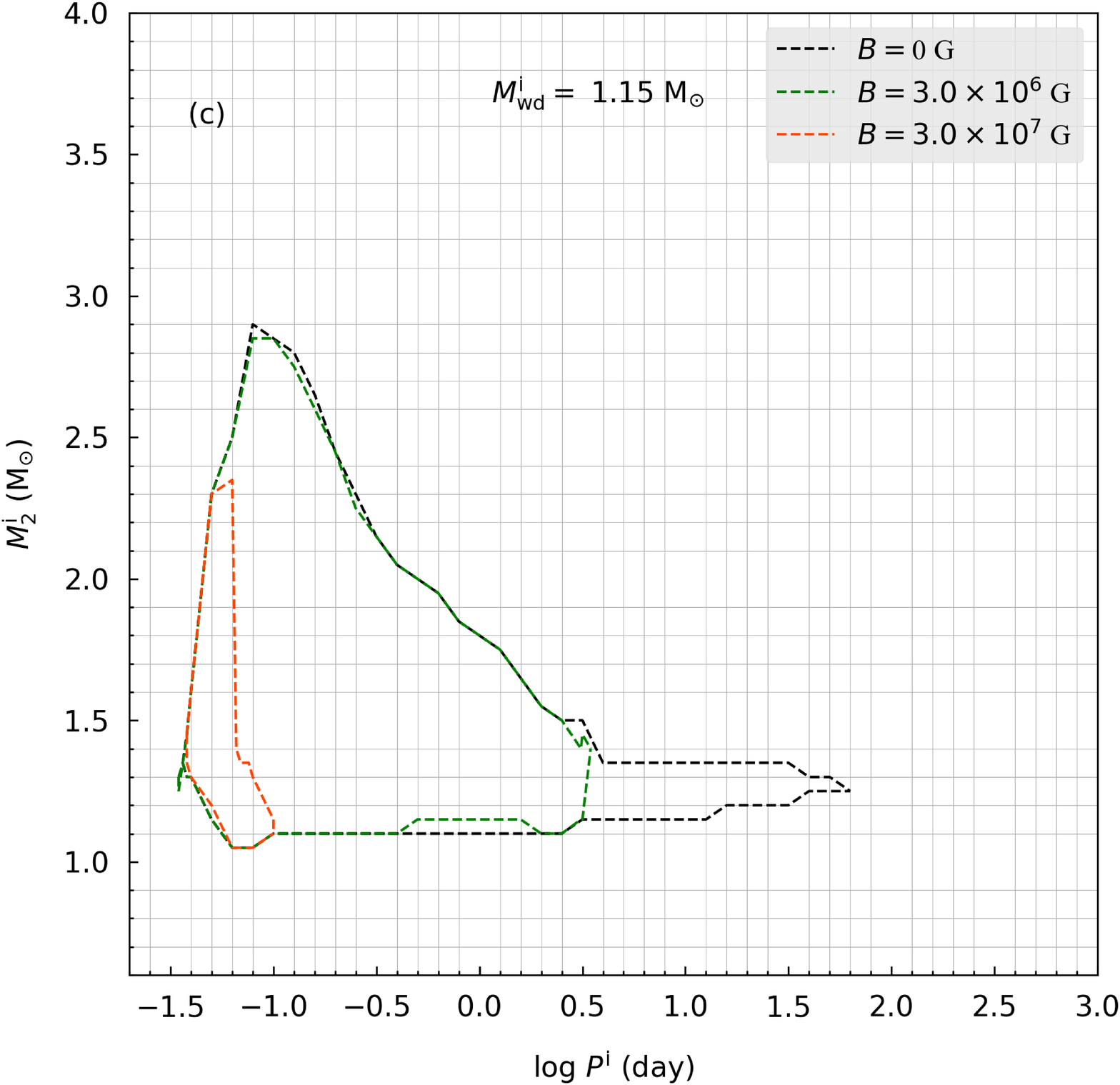}
	\includegraphics[width=0.48\linewidth]{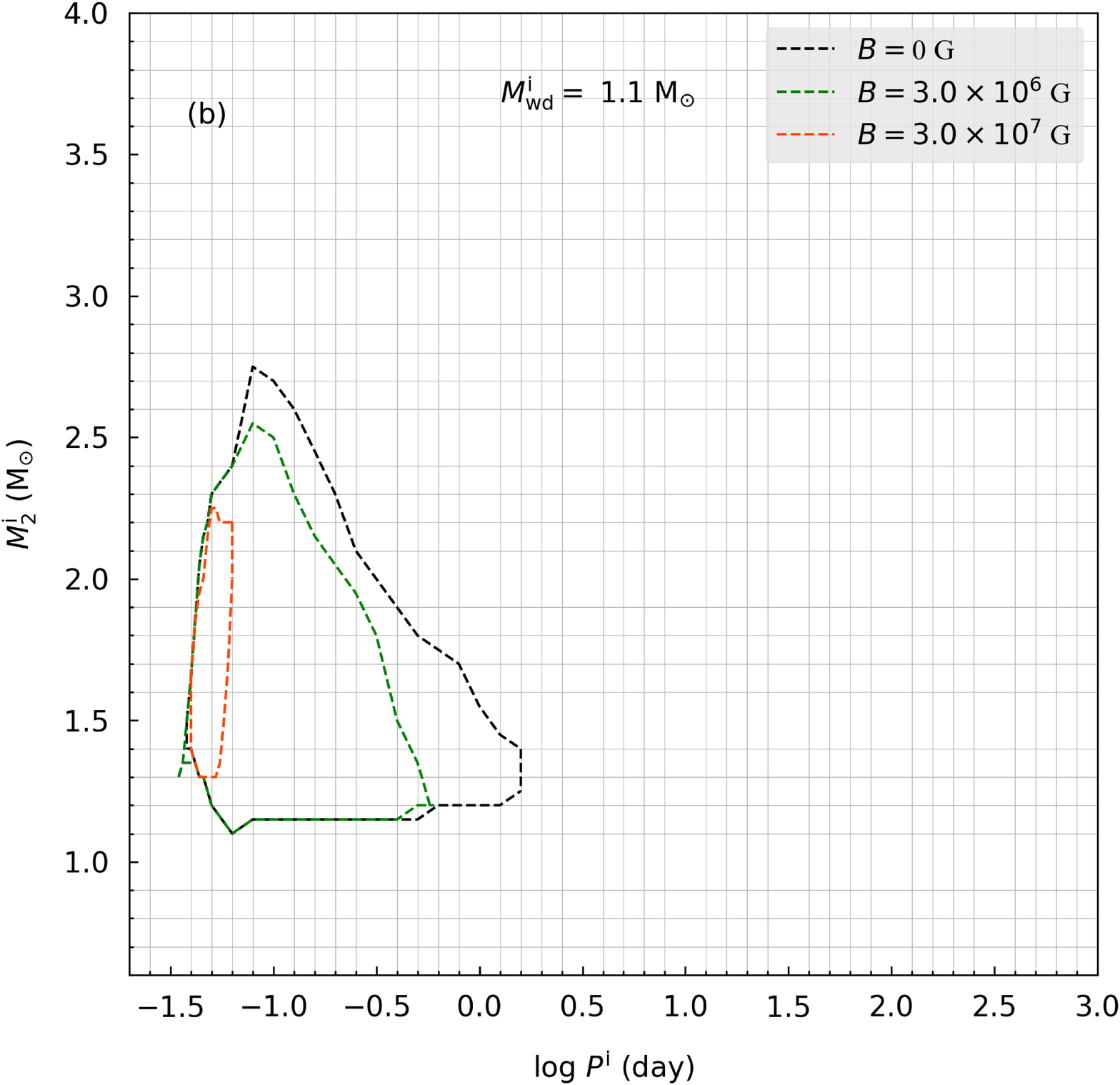}
	\includegraphics[width=0.48\linewidth]{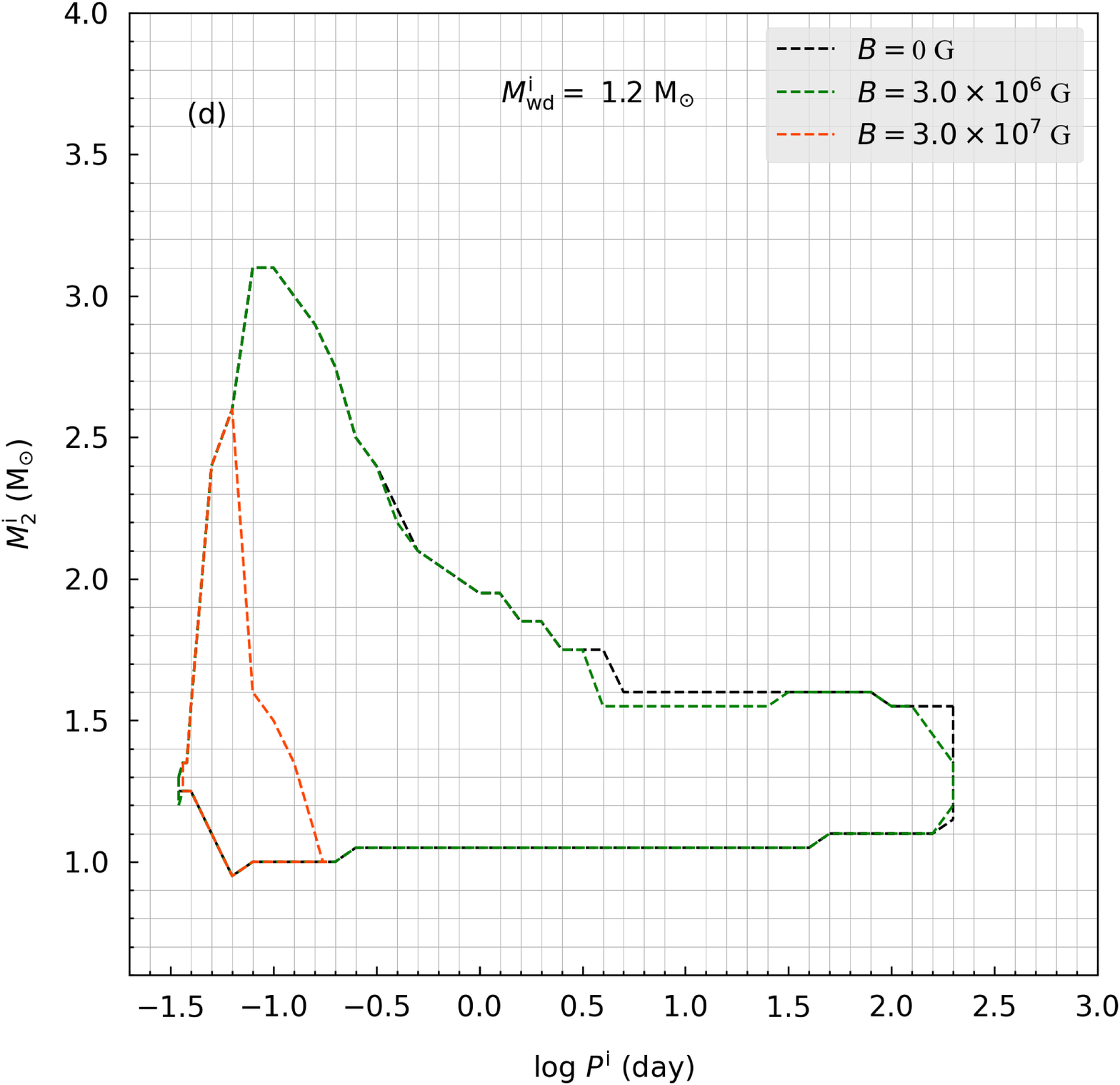}	
	\caption{Distribution of the initial orbital period $P^{\rm i}$ and companion mass $M_{\rm 2}^{\rm i}$ for potential SN Ia progenitor systems with different $M_{\rm wd}^{\rm i}$ and $B$-fields. The KH prescription is adopted for the He-rich mass accumulation efficiency. The four panels correspond to $M_{\rm wd}^{\rm i} = 1.0 \rm M_\odot$ (panel [a]), $1.1 \rm M_\odot$ (panel [b]), $1.15 \rm M_\odot$ (panel [c]) and $1.2 \rm M_\odot$ (panel [d]), respectively. The black, green and orange-red curves represent the results with $B = 0$, $3.0 \times 10^6$ and $3.0 \times 10^7$ G, respectively. \label{fig:f4}}
\end{figure}

\begin{figure}
	\centering
	\includegraphics[width=0.48\linewidth]{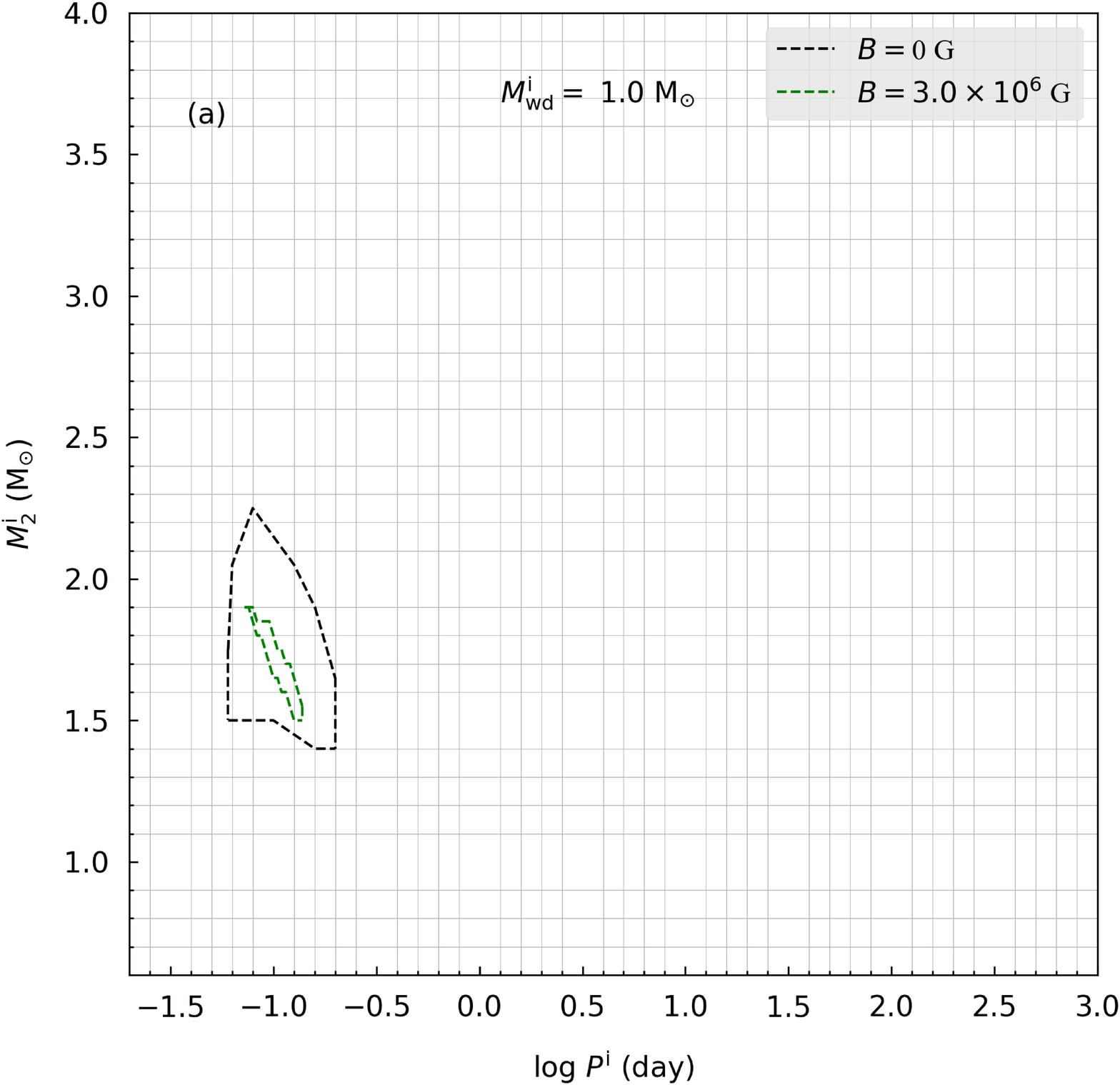}
	\includegraphics[width=0.48\linewidth]{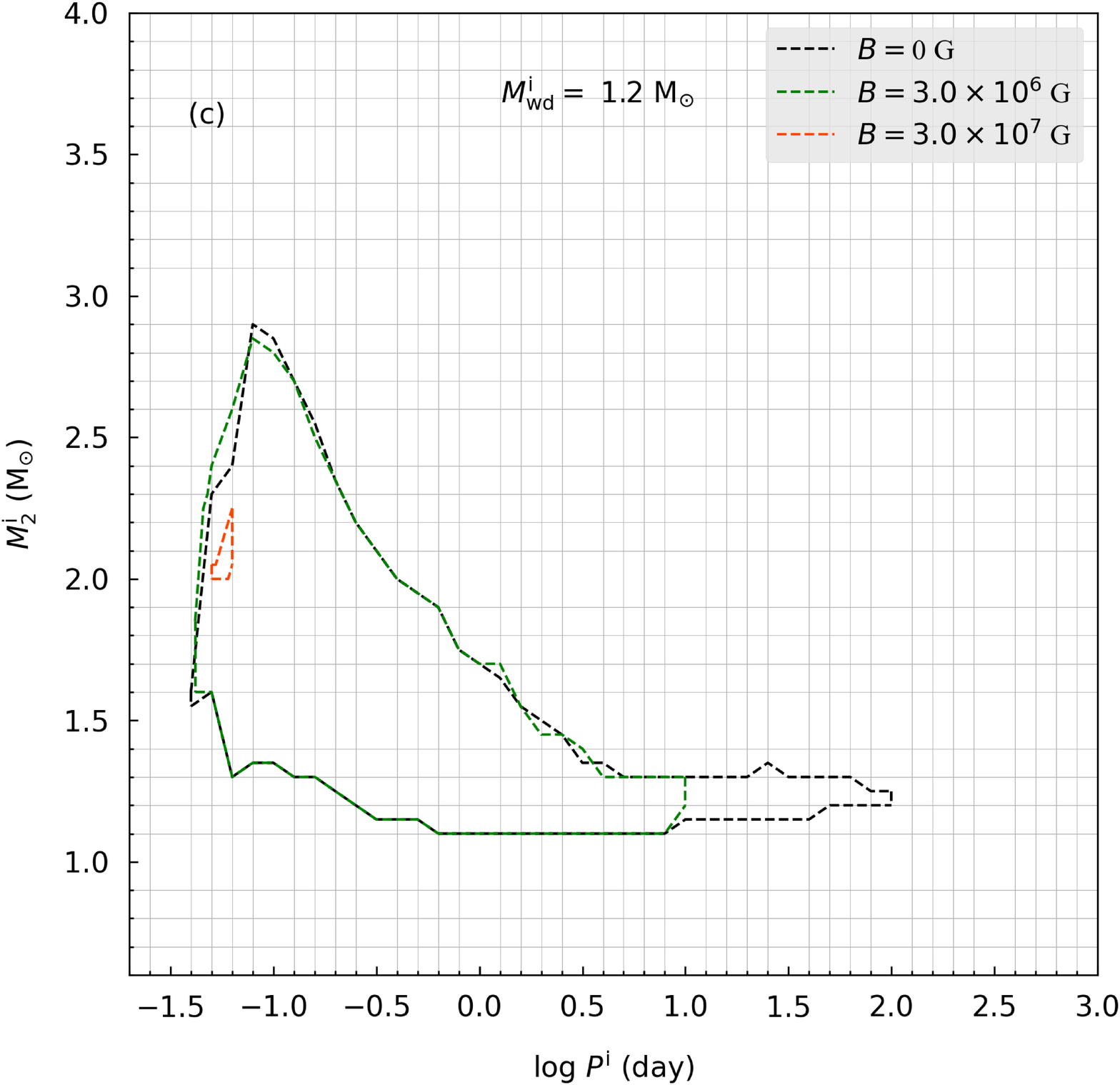}
	\includegraphics[width=0.48\linewidth]{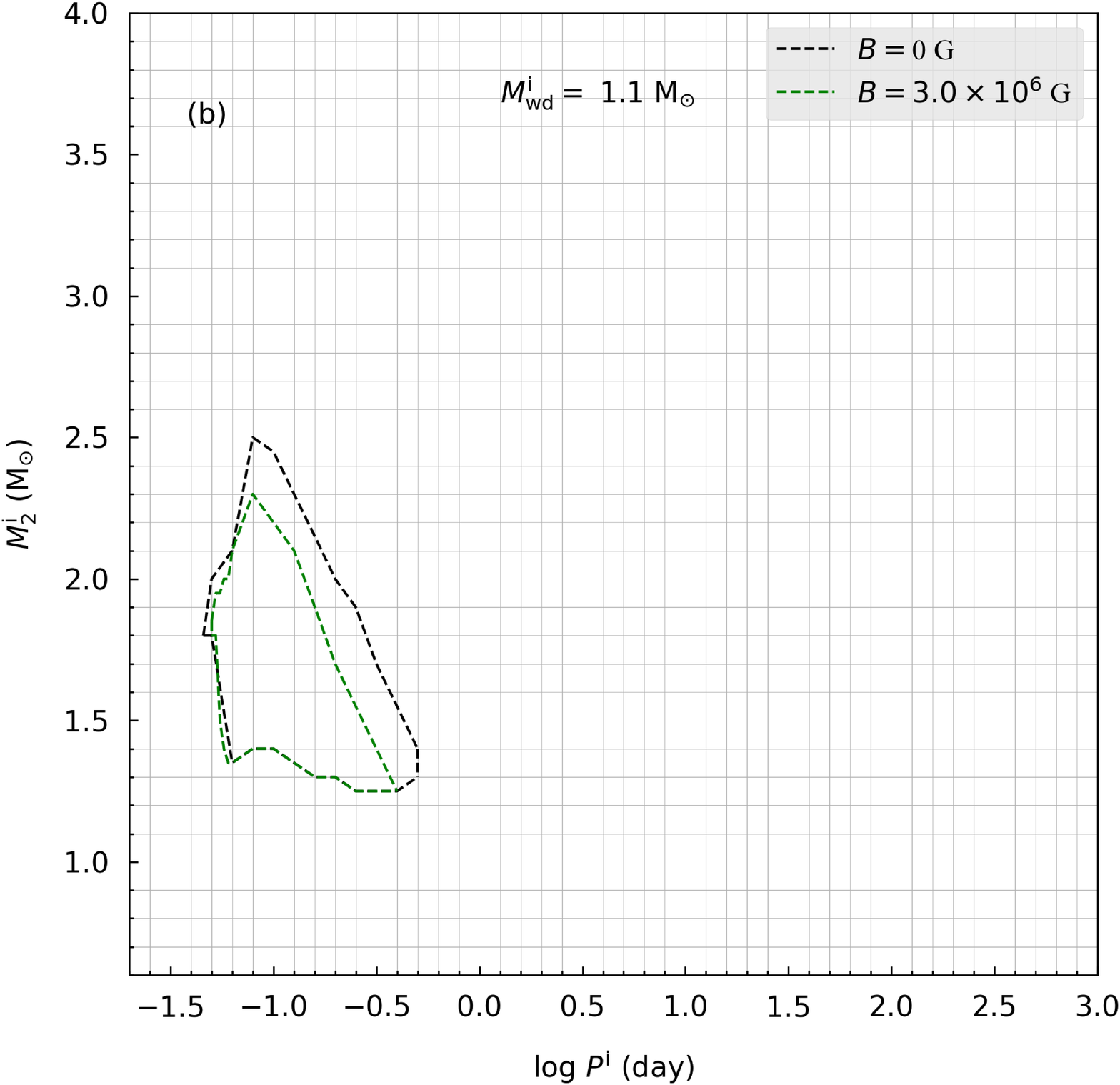}
	\includegraphics[width=0.48\linewidth]{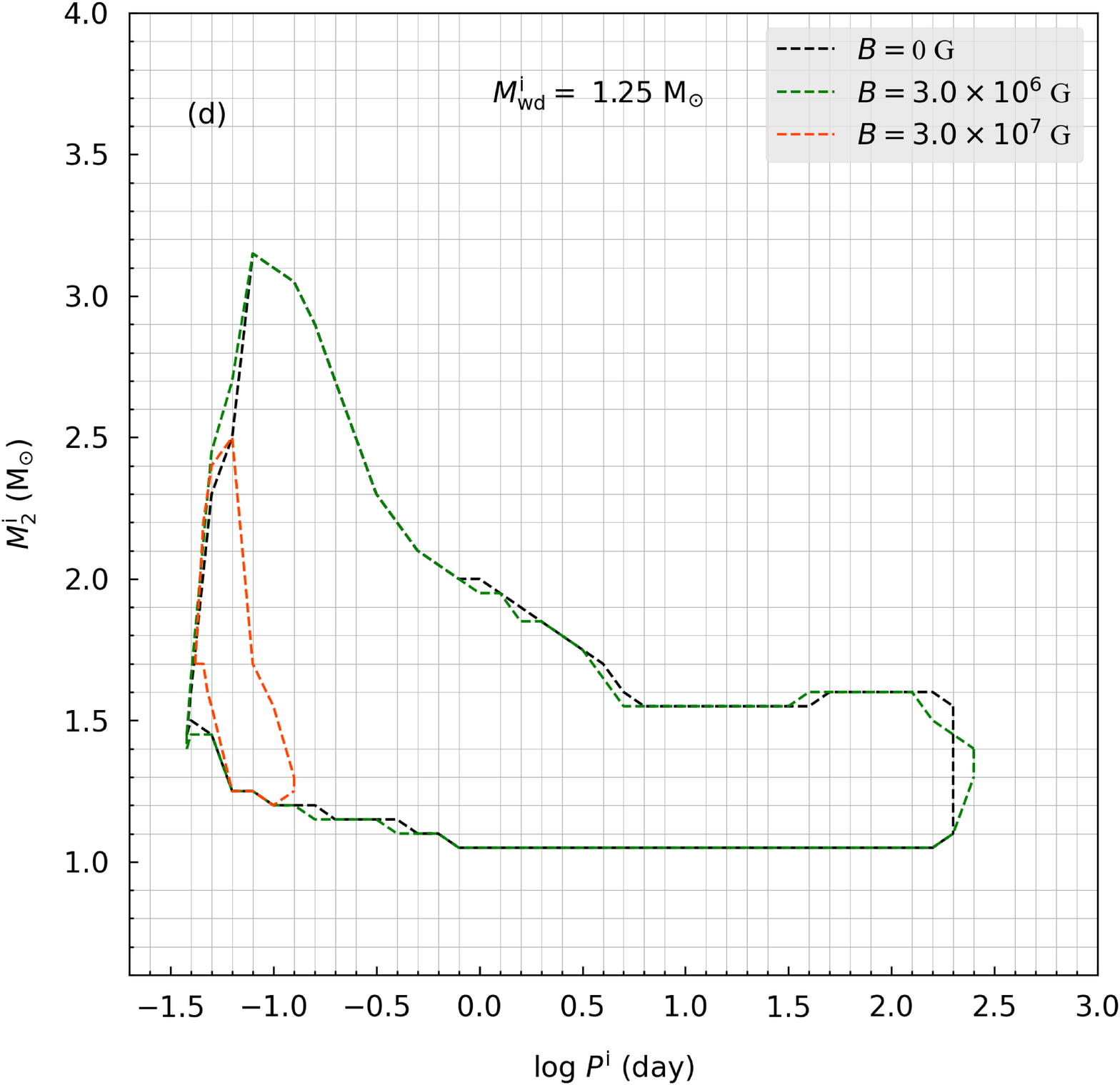}
	\caption{Same as Fig.~\ref{fig:f4} but with the WL prescription adopted for the He-rich mass accumulation efficiency. Panels [a]-[d] correspond to $M_{\rm wd}^{\rm i}$ = $1.0$, $1.1$, $1.2$ and $1.25$ $\rm M_\odot$, respectively. Note that the minimal initial mass of WD that can form SN Ia is 1.2 $\rm M_{\odot}$ in the high-magnetic model.  \label{fig:f5}}
\end{figure}

Fig.~\ref{fig:f4} shows the initial distributions of the SN Ia progenitors on the $P^{\rm i}-M_{\rm 2}^{\rm i}$ plane with different $B$-fields. The four panels correspond to $M_{\rm wd}^{\rm i}=$ 1.0, 1.1, 1.15 and 1.2 $M_\odot$ (the minimum $M_{\rm wd}^{\rm i}$ that can form SN Ia is 0.858, 0.95 and 1.0 $\rm M_{\odot}$ in the non-, intermediate- and high-magnetic models, respectively), respectively. It is obvious that the parameter spaces shrink toward shorter orbital periods and smaller donor masses with increasing $B$, and for a given magnetic field strength, the parameter spaces become larger for more massive $M_{\rm wd}^{\rm i}$. We also show the corresponding results with the WL prescription in Fig.~\ref{fig:f5}. Since the stable He-burning region obtained by \cite{wang2015} is substantially narrower than by \cite{nomoto1982}, the initial parameter space for the progenitors of SNe Ia with the WL prescription is considerably smaller than with the KH prescription. Only WDs more massive than 1.2 $\rm M_{\odot}$ can evolve to SNe Ia if their magnetic fields are as high as $3\times 10^7$ G. These mean that the magnetic fields of WDs do not help stabilize helium burning on WDs, in contrast with the results for the hydrogen burning processes. To investigate their influence on the overall SN Ia production in our Galaxy, we perform BPS of the evolution of magnetized WDs and present the results in the next section.

\section{Binary population synthesis of WD + He binary evolution} \label{subsec:bps}
We use the BPS code originally developed by \citet{hurley2002} to simulate the evolution of $2\times 10^7$ binaries until the formation of WD + He star binaries. Theoretically, the WD + He star binaries can form in three major ways \citep{wang2012},
\begin{enumerate}
	\item $\rm{MS}\;+\;\rm{MS}{\longrightarrow}\rm{subgaint}/\rm{FGB}\;+\;\rm{MS}\stackrel{\rm{stabel}\;\rm{RLOF}}{\longrightarrow}\rm{He}\;\rm{star}+ \;\rm{MS}\stackrel{\rm{stable}\;\rm{RLOF}}{\longrightarrow}\rm{CO}\;\rm{WD} \;+\; \rm{MS}\stackrel{\rm{dynamical}\;\rm{unstable}\;\rm{RLOF}}{\longrightarrow}\rm{CE}\;\rm{phase}\stackrel{\rm{CE} \; \rm{ejection}}{\longrightarrow}\rm{CO} \; \rm{WD} \;+ \;\rm{He} \; \rm{star}$,
	\item $\rm{MS}\;+\;\rm{MS}{\longrightarrow}\rm{EAGB}\;\rm{star}\;+\;\rm{MS}\stackrel{\rm{dynamical}\;\rm{unstabel}\;\rm{RLOF}}{\longrightarrow}\rm{CE}\;\rm{phase}\stackrel{\rm{CE}\; \rm{ejection}}{\longrightarrow}\rm{He}\;\rm{RG}\;+ \;\rm{MS}\stackrel{\rm{stable}\;\rm{RLOF}}{\longrightarrow}\rm{CO}\;\rm{WD}\;+ \;\rm{MS}\stackrel{\rm{dynamical}\;\rm{unstable}\;\rm{RLOF}}{\longrightarrow}\rm{CE}\;\rm{phase}\stackrel{\rm{CE} \; \rm{ejection}}{\longrightarrow}\rm{CO} \; \rm{WD} \;+ \;\rm{He} \; \rm{star}$,
	\item $\rm{MS}\;+\;\rm{MS}{\longrightarrow}\rm{TPAGB}\;\rm{stage}\;+\;\rm{He-core}\;\rm{burning}\;\rm{stage}\stackrel{\rm{dynamical}\;\rm{unstable}\;\rm{RLOF}}{\longrightarrow}\rm{CE}\;\rm{phase}\stackrel{\rm{CE} \; \rm{ejection}}{\longrightarrow}\rm{CO} \; \rm{WD} \;+ \;\rm{He} \;\rm{star}$.
\end{enumerate}
Here $\rm{FGB}$, $\rm{EAGB}$ and $\rm{TPAGB}$ are the abbreviations of the stellar evolutionary stage of first giant branch, early asymptotic giant branch, and thermally pulsating asymptotic giant branch, respectively. We then combine the properties of the WD + He star binaries from our BPS calculations with the subsequent evolutionary results calculated with the MESA code. If the parameters of the formed CO WD + He star systems are located in the parameter space shown in Fig.~\ref{fig:f4} or Fig.~\ref{fig:f5}, we assume that it will explode as an SN Ia. From the calculated results we can obtain the birthrates and delay times of the SNe Ia, and the characteristics of the surviving companion stars.

In the BPS study, the initial primary star mass $M_1$, secondary star mass $M_2$ and orbital separations $a$ are set to be $0.8 - 40 \rm M_\odot$, $0.8 - 40 \rm M_\odot$ and $3 - 10^4 R_\odot$, respectively. We assume that the distribution of $M_1$ follows the initial mass function of \cite{kroupa1993}, the mass ratio $q_2 =  M_2/M_1$ is uniformly distributed within $[0,1]$, and the distribution of ln $a$ is also uniformly distributed. Same as \citet{hurley2002}, we assume one binary systems with $M_1 \geq 0.8 \rm M_\odot$ is born in the Galaxy per year, i.e. the star formation rate is 7.6085 $\rm yr^{-1}$. When the mass transfer is dynamically unstable, a common envelope (CE) phase ensues, and the accreting star spirals into the donor's envelope. To deal with the CE evolution, we adopt the values of the binding energy parameter $\lambda$ provided by \citet{xu2010} for the envelope of the donor, and the efficiency factor $\alpha =0.5$ and 1.0. All the binaries are assumed to be in circular orbits.

\begin{figure}[htp]
	\centering
	\includegraphics[width=0.33\linewidth]{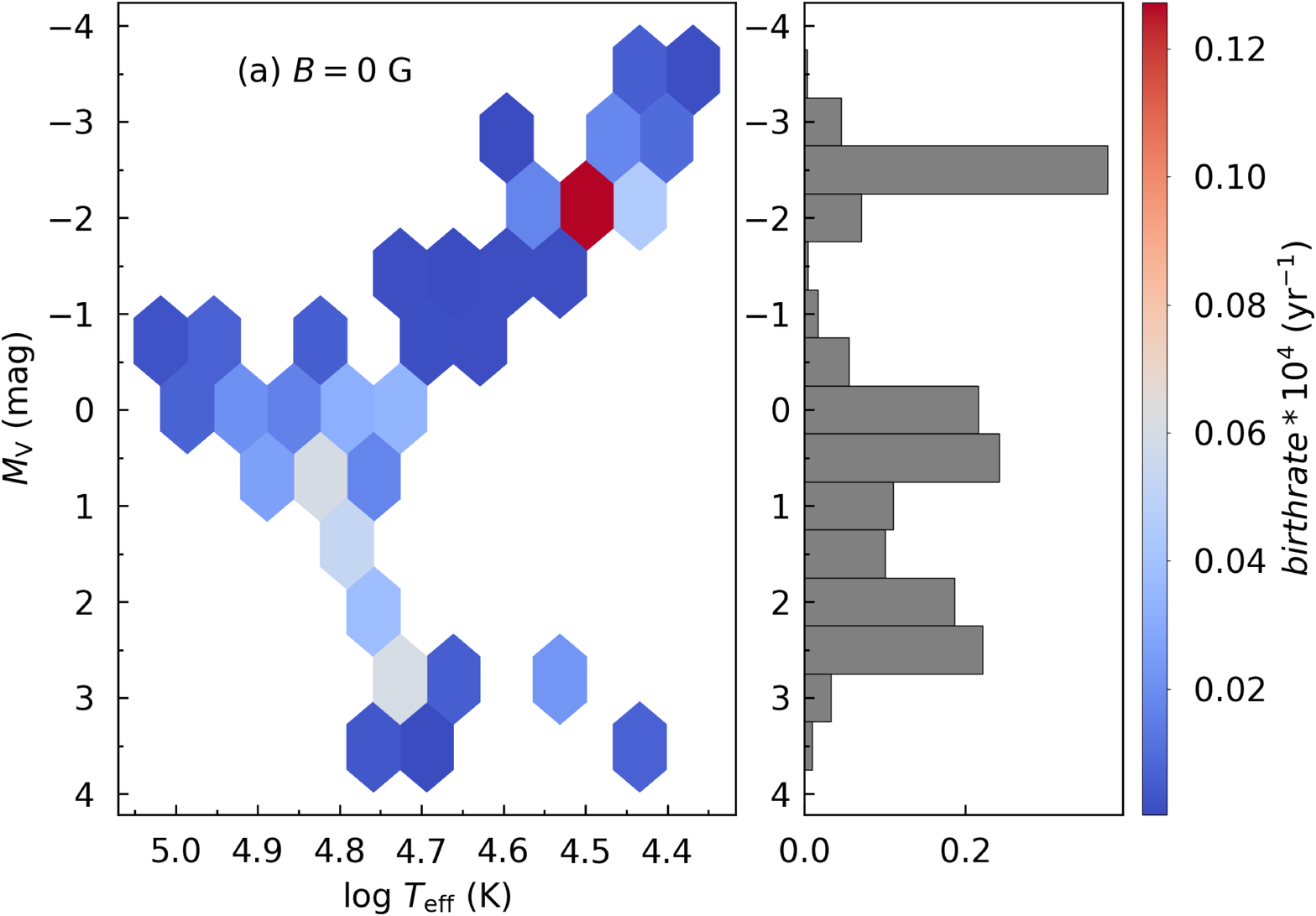}
	\includegraphics[width=0.33\linewidth]{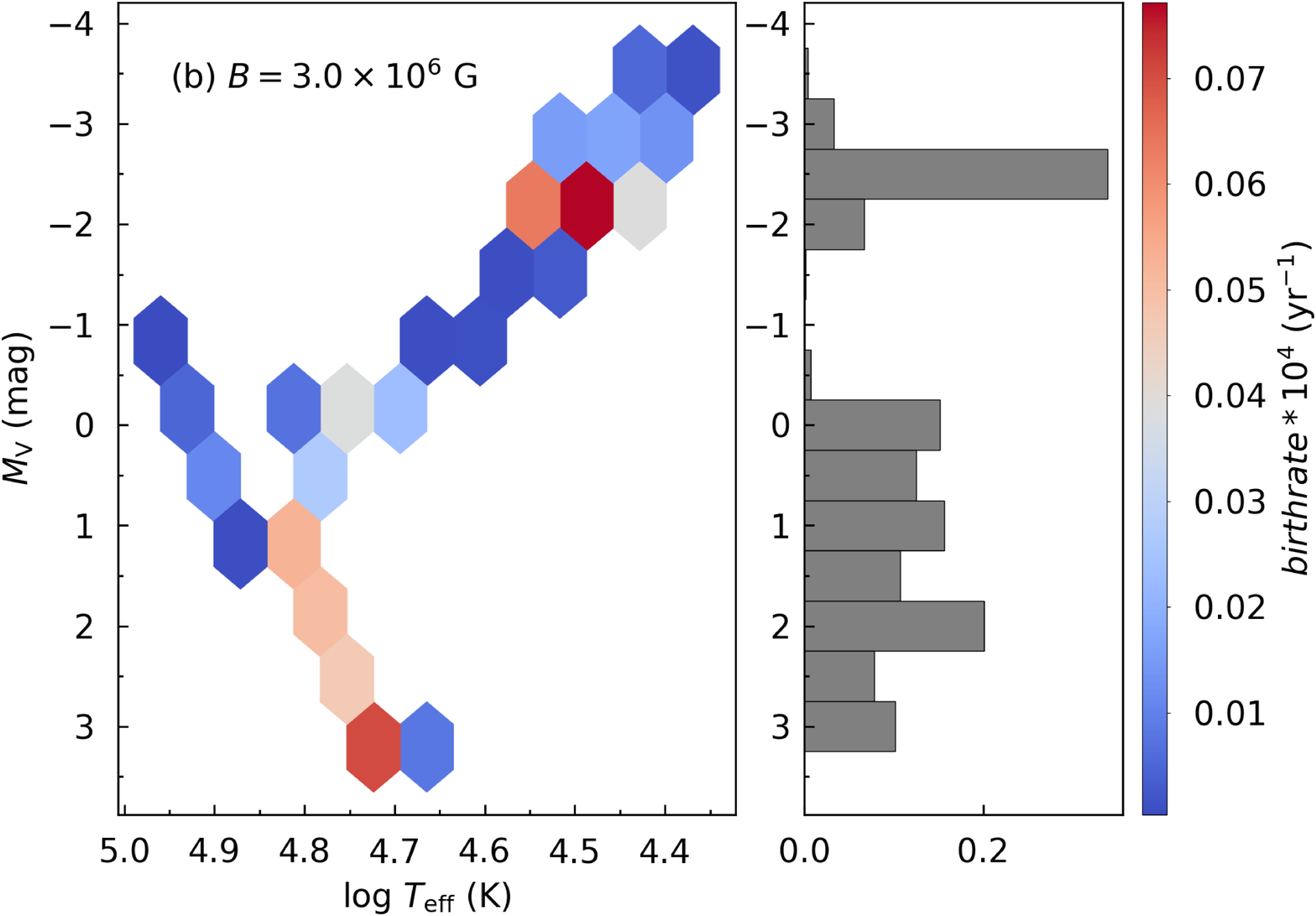}
	\includegraphics[width=0.33\linewidth]{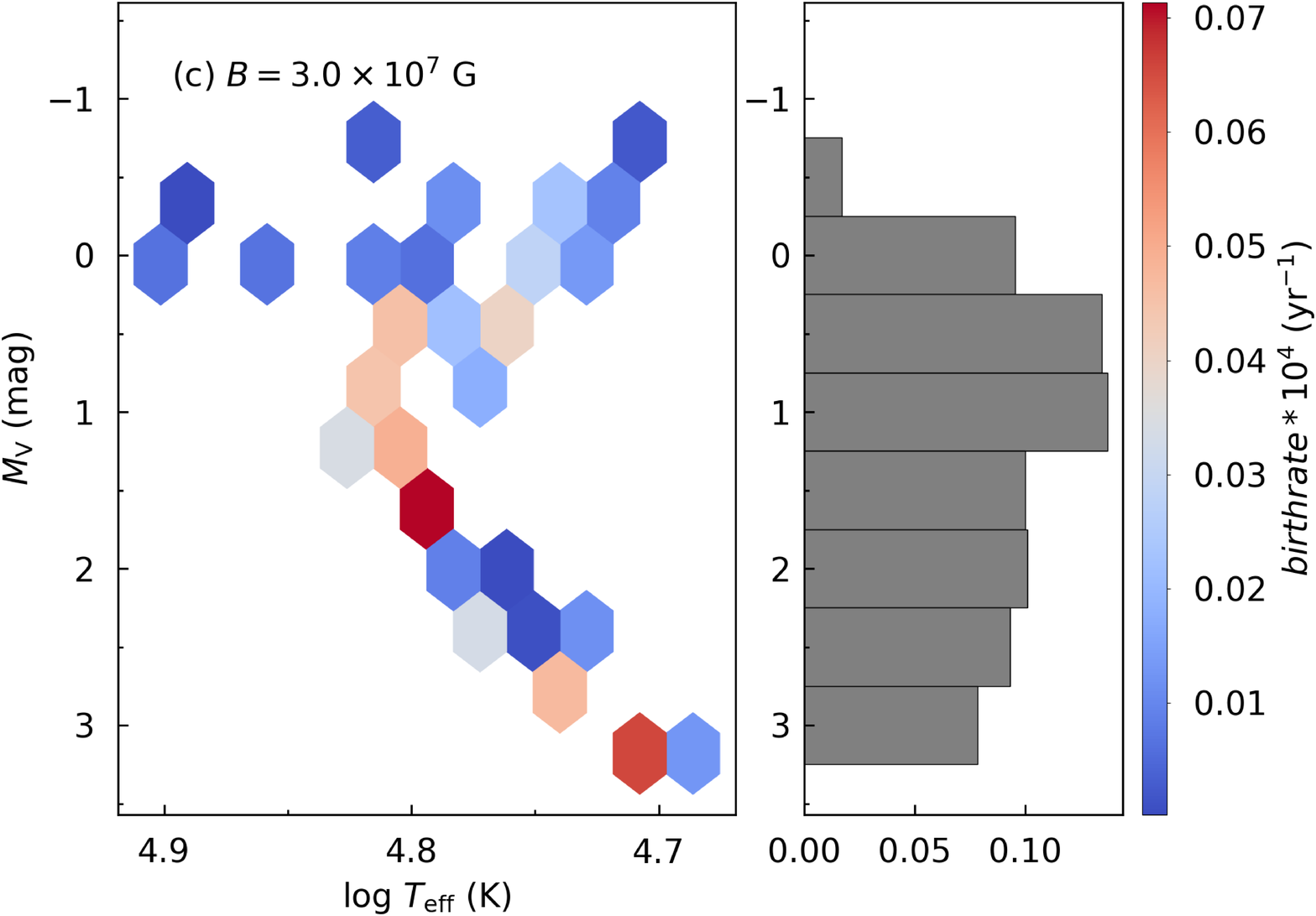}
	\caption{Distribution of the effective temperature and the absolute magnitude of the surviving companion at the moment of SN explosion with $\alpha=0.5$. Panels from left to right show the outcomes of $B = 0$, $3.0 \times 10^6$ and $3.0 \times 10^7$ G, respectively. The horizontal histogram close to the right of each panel gives the birthrate distribution of $M_{\rm V}$. The KH prescription is adopted for the He-rich mass accumulation efficiency. The colors of the hexagons indicate the birthrates of these systems. \label{fig:f6}}
\end{figure}

\begin{figure}[htp]
	\centering
	\includegraphics[width=0.33\linewidth]{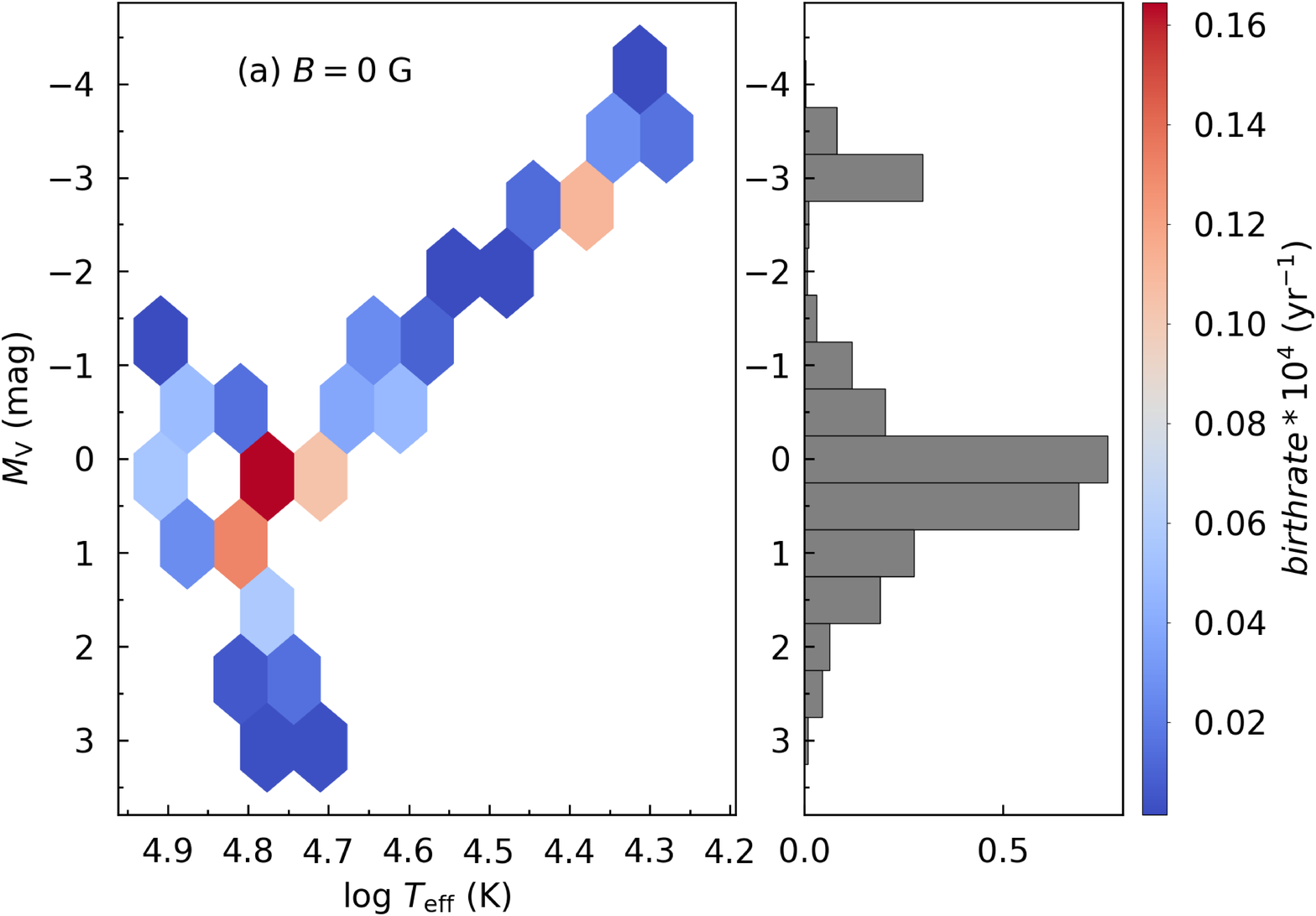}
	\includegraphics[width=0.33\linewidth]{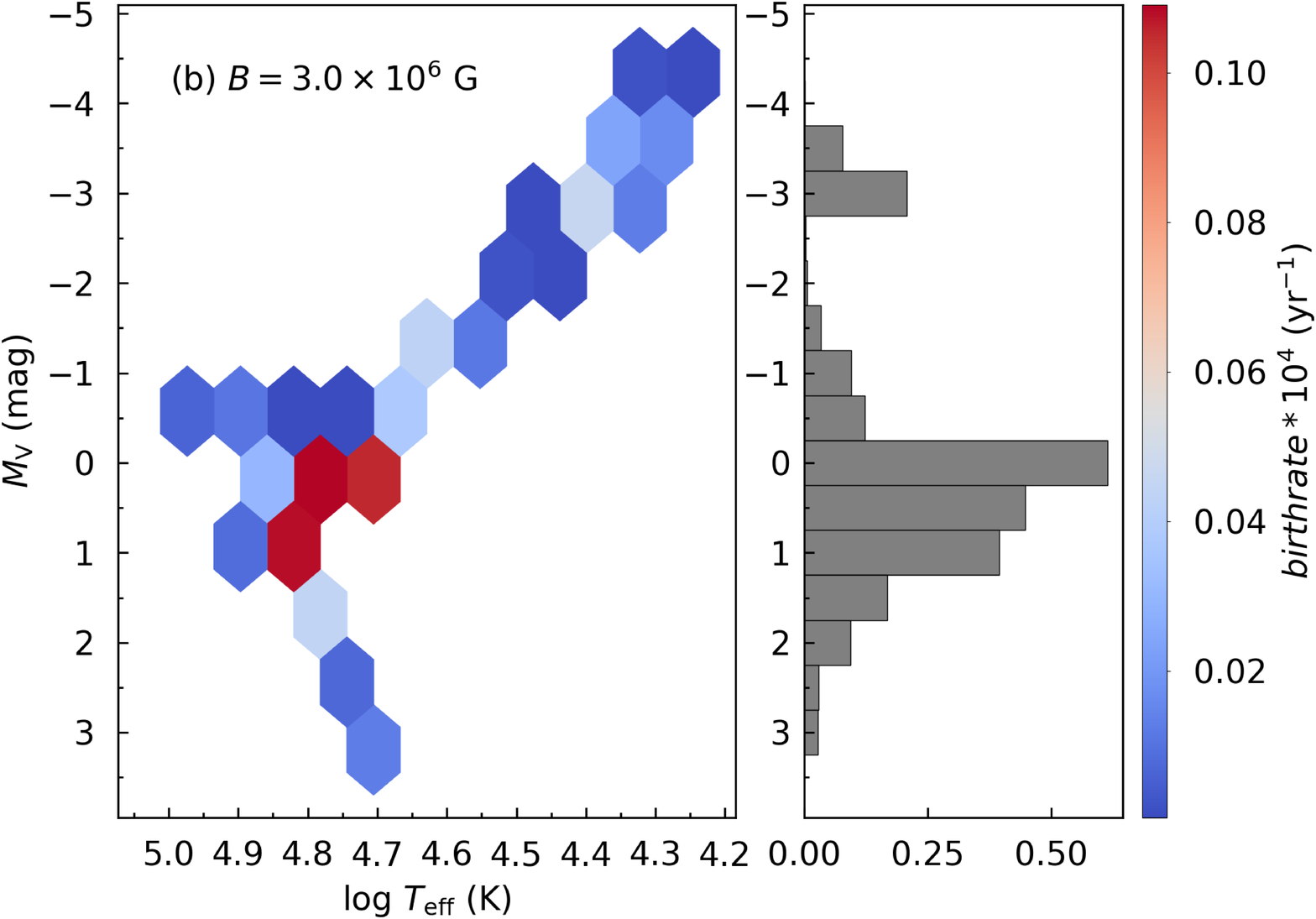}
	\includegraphics[width=0.33\linewidth]{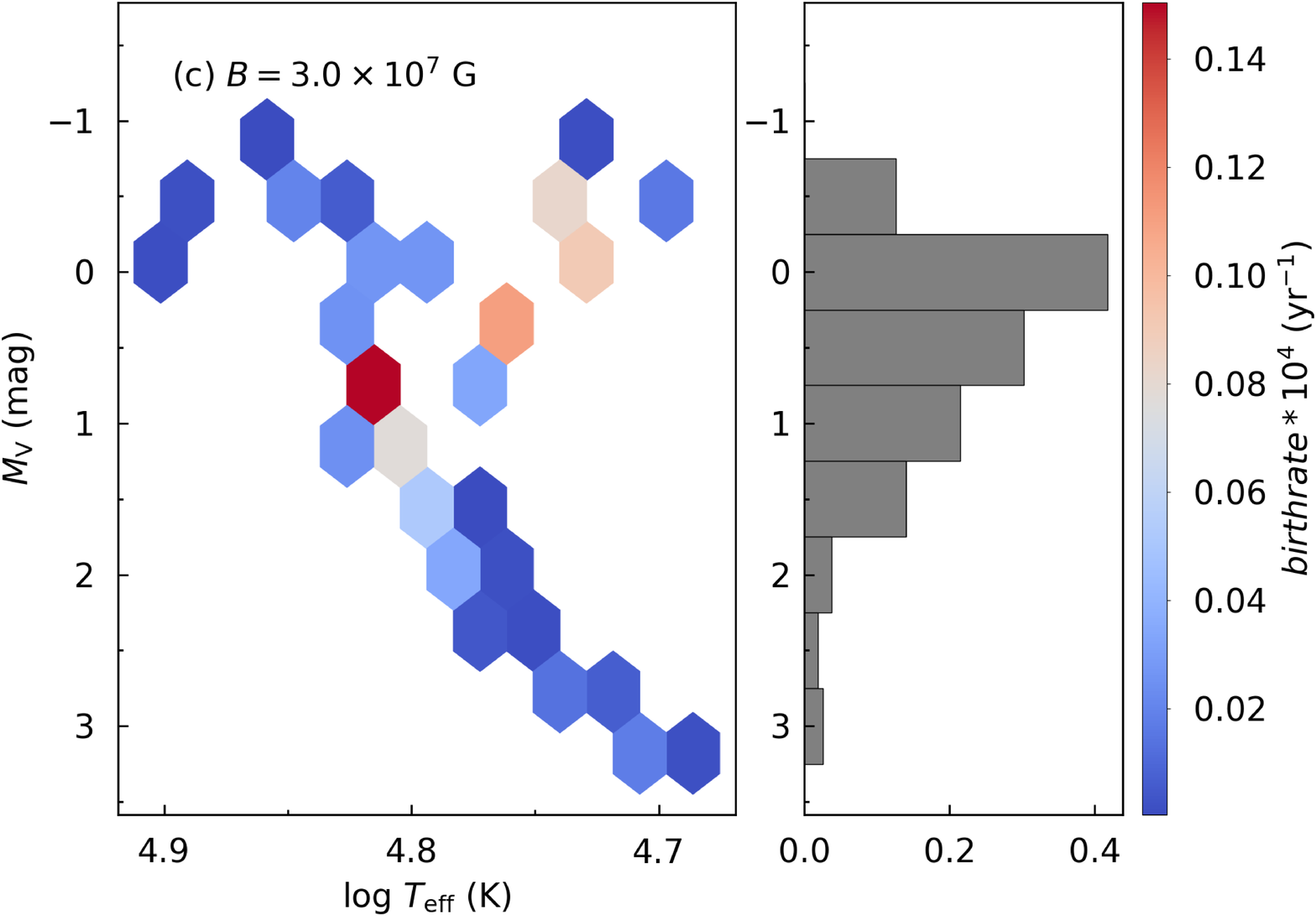}
	\caption{Same as Fig.~\ref{fig:f6} but with $\alpha = 1.0$. \label{fig:f7}}
\end{figure}

Figs.~\ref{fig:f6} and ~\ref{fig:f7} show the companion star's absolute magnitude in V-band ($M_{\rm V}$) and effective temperature ($T_{\rm eff}$) at the moment of the SN explosion, with $\alpha = 0.5$ and 1.0 respectively. Here the He-rich mass accumulation efficiencies are calculated with the KH prescription. Each panel demonstrates the results with a specific magnetic field strength, and the colors denote the magnitude of the SN birthrate. Next to each panel we show the histogram distribution of $M_{\rm V}$. Both the non- and intermediate-magnetic models demonstrate similar bimodal distributions (the bright peak is provided by the more massive donors in wide systems and the dim peak denotes the less massive ones in compact systems), while the distribution in the high-magnetic model is considerably narrower than in the former two models. A large fraction of the surviving companion stars are brighter than $\sim -2.5^m$ when $B = 0$ and $3.0 \times 10^6$ G, but mainly dimmer than $\sim 0^m$ when $B = 3.0 \times 10^7$ G. Fig.~\ref{fig:f7} also shows a bimodal $M_{\rm V}$ distribution, peaking at $M_{\rm V}\sim -3^{m}$ and $0^m$ in both the non- and intermediate-magnetic models, while the $M_{\rm V}$ distribution is clustered around $\sim 0^m$ in the high-magnetic model.  Roughly speaking, the stronger the magnetic fields, the dimmer the surviving companions. One apparent difference between the results with $\alpha=0.5$ and $\alpha=1.0$ is that there are more surviving companions with $M_{\rm V} \sim 0^m-1^m$ in the latter. They are the survivors of compact progenitor systems composed of a less massive He star, which more likely merge during the CE phase with $\alpha=0.5$. The brightness limits of the SN 2011fe's progenitor system make it impossible for the companion with mass $\geq 3.5 \rm M_\odot$ and $M_{\rm V}\leq 1^m$, and rule out almost all the bright companions predicted from the RG-donor and He-donor channels \citep{li2011}. However, quite a part of the companions in our high-magnetic models reveal $M_{\rm V}\geq 0^m$, which are still compatible with the observations of SN 2011fe.

\begin{figure}[htp]
	\centering
	\includegraphics[width=0.33\linewidth]{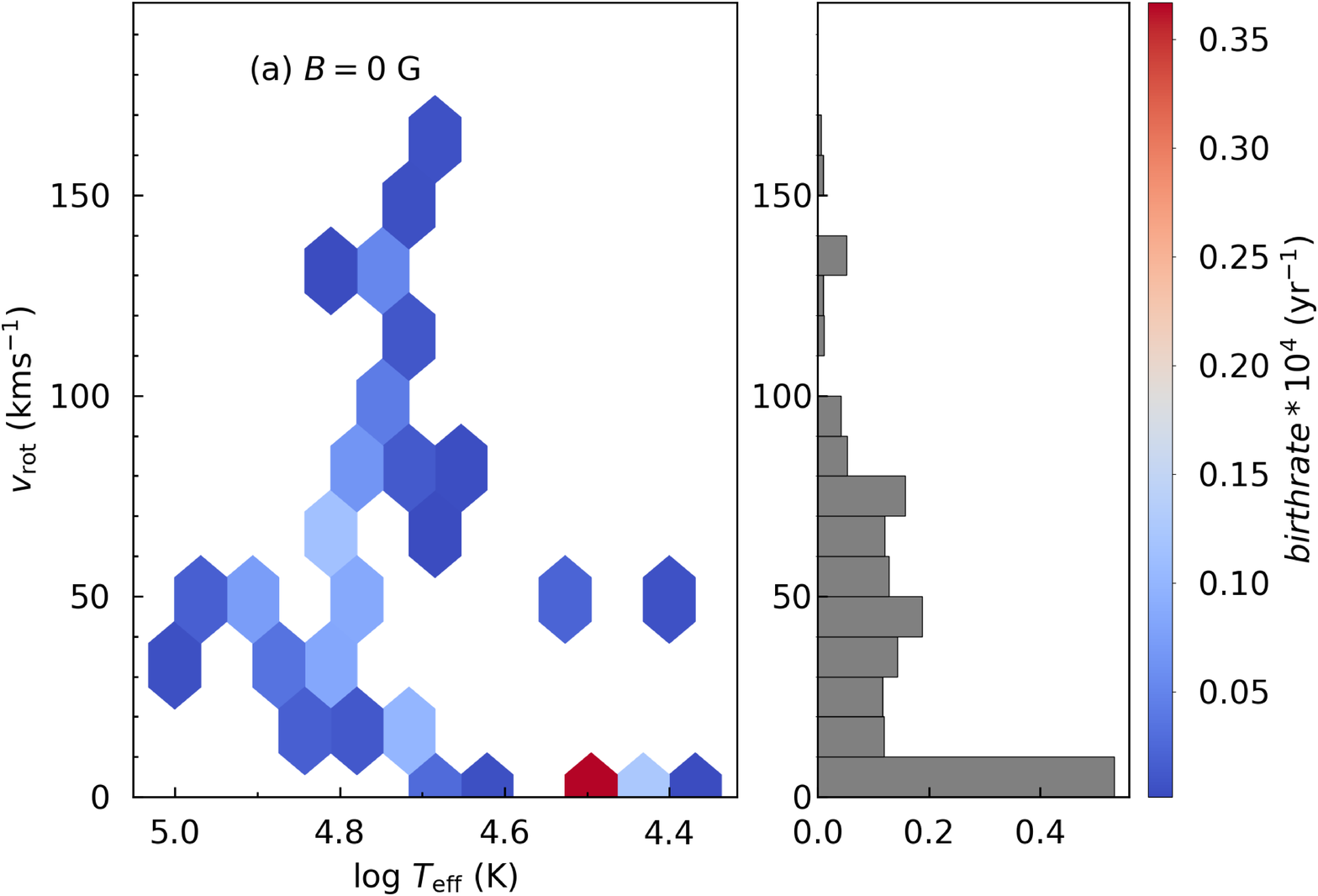}
	\includegraphics[width=0.33\linewidth]{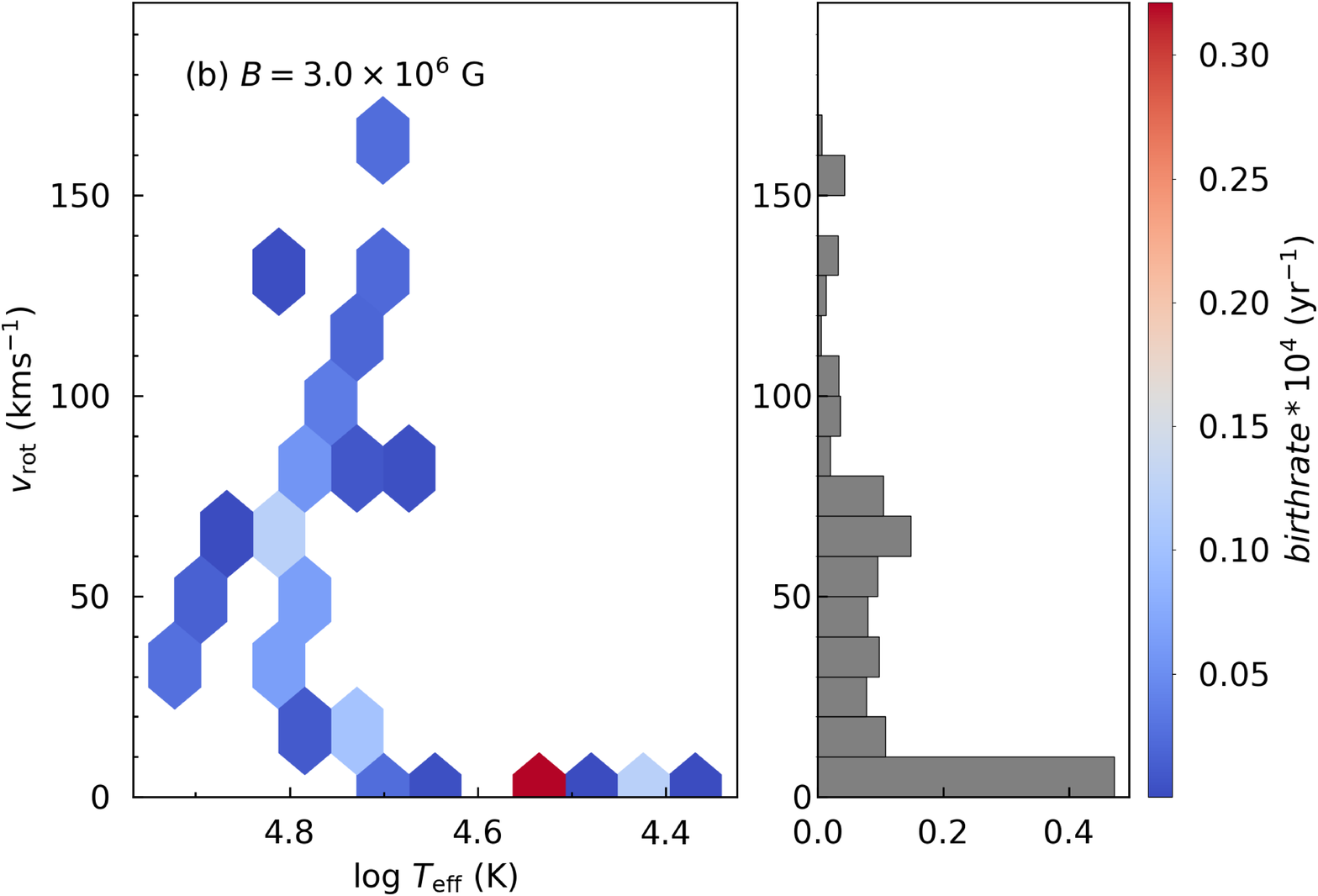}
	\includegraphics[width=0.33\linewidth]{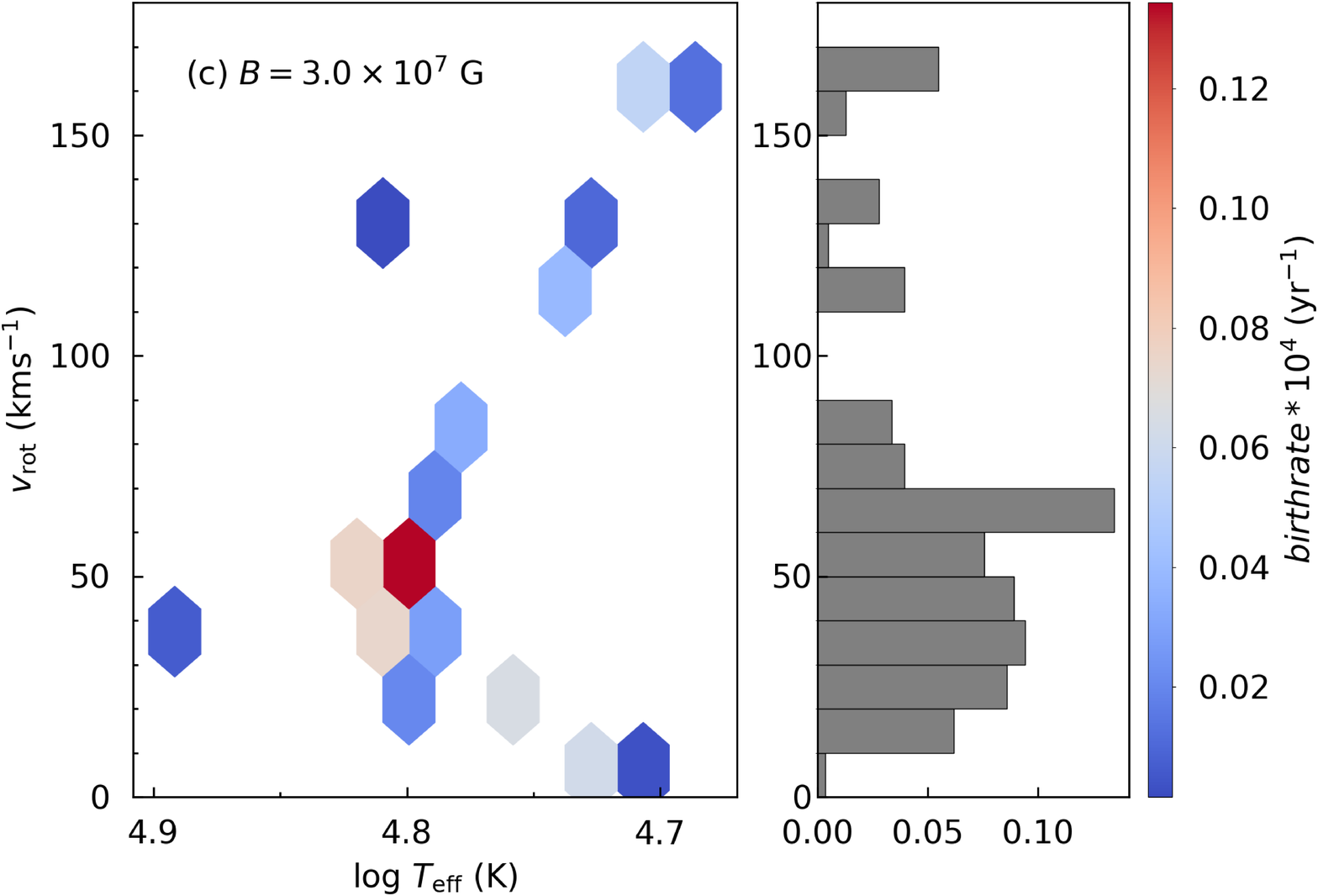}
	\caption{Distribution of the surface rotational velocity and effective temperature of the surviving companions at the moment of SN explosion with $\alpha=0.5$. Panels from left to right show the outcomes with $B = 0$, $3.0 \times 10^6$ and $3.0 \times 10^7$ G, respectively. Birthrate distribution of $v_{\rm rot}$ is also shown in the corresponding horizontal histogram. The colors of the hexagons indicate the corresponding birthrates of these systems. \label{fig:f8}}
\end{figure}

\begin{figure}[htp]
	\centering
	\includegraphics[width=0.33\linewidth]{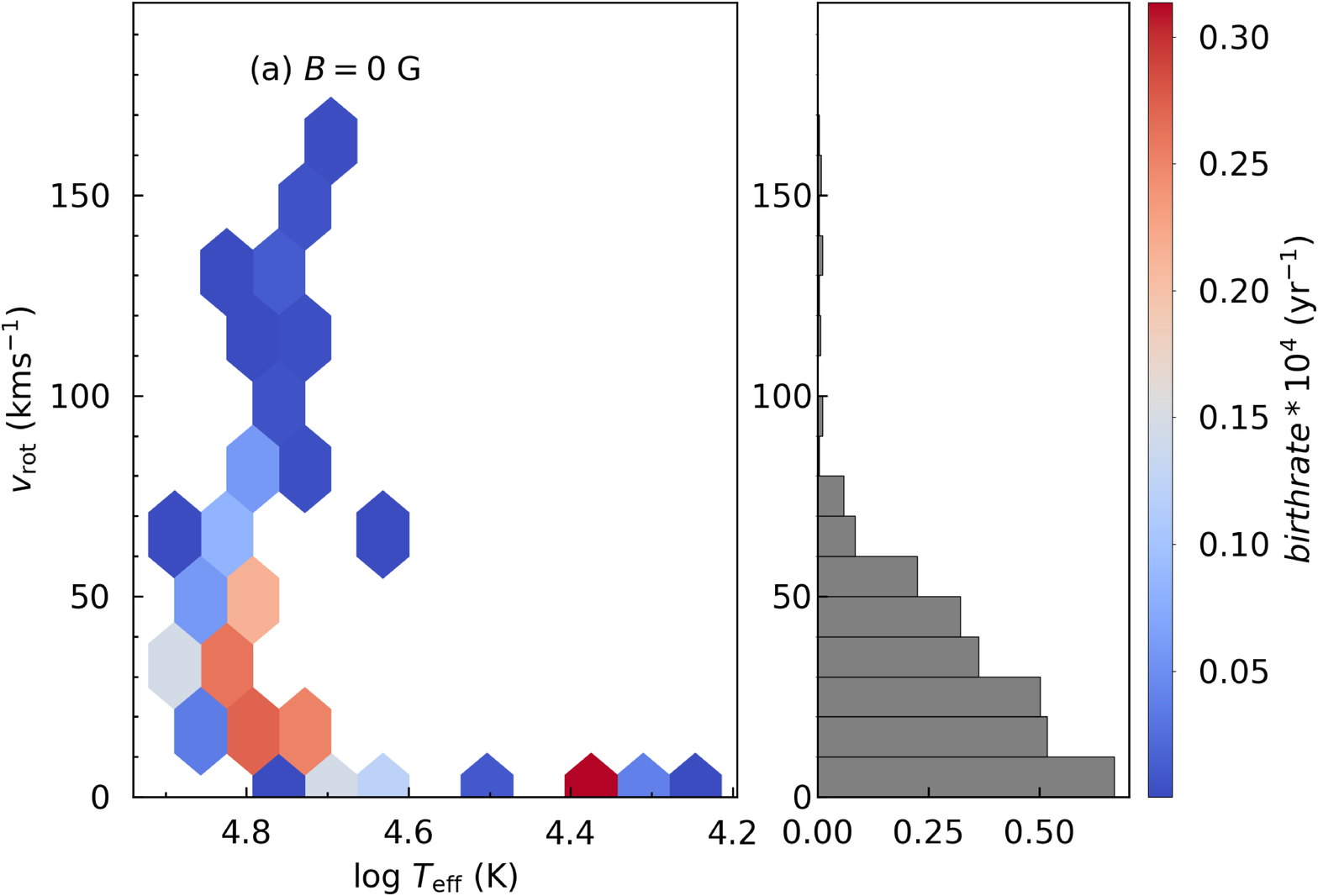}
	\includegraphics[width=0.33\linewidth]{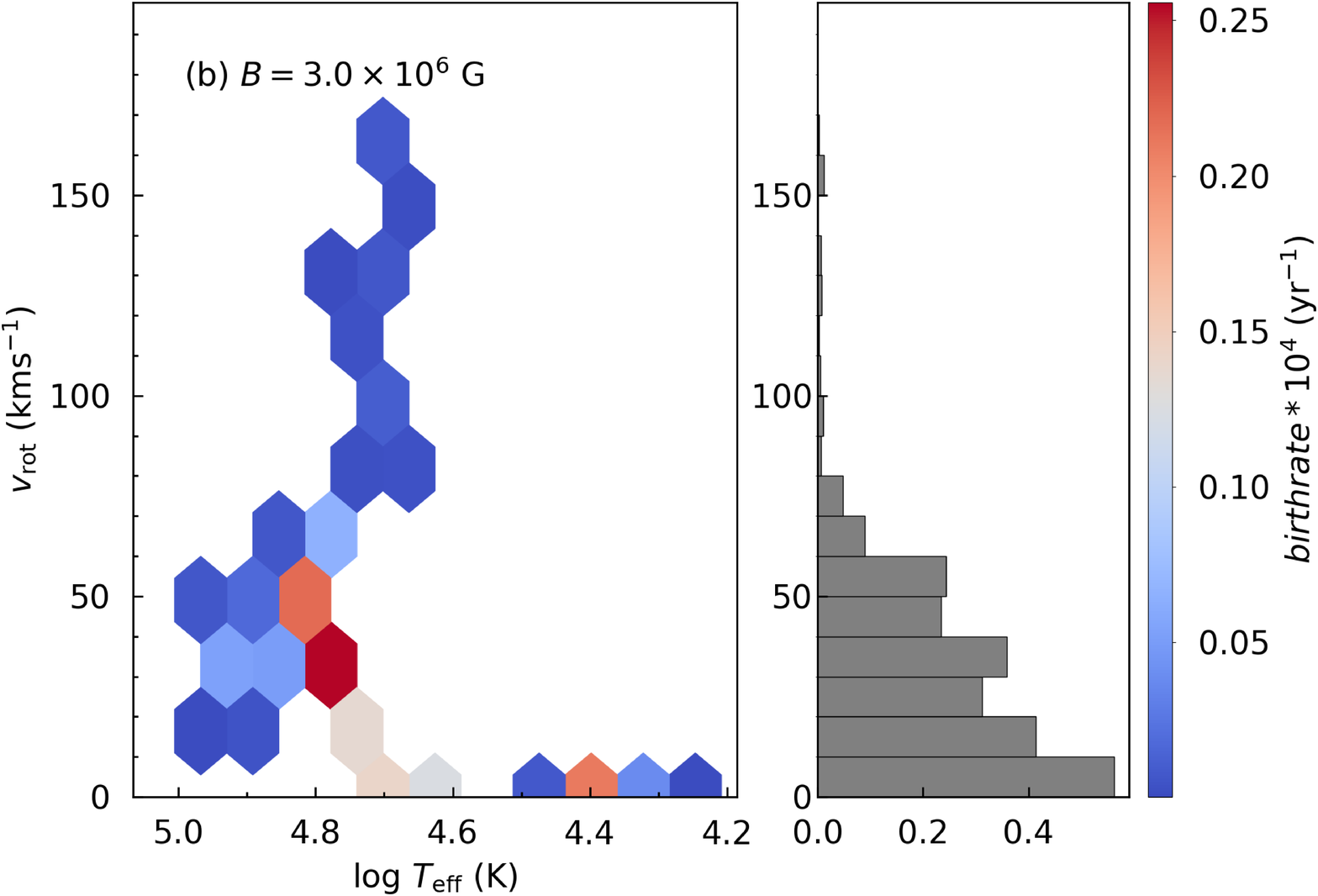}
	\includegraphics[width=0.33\linewidth]{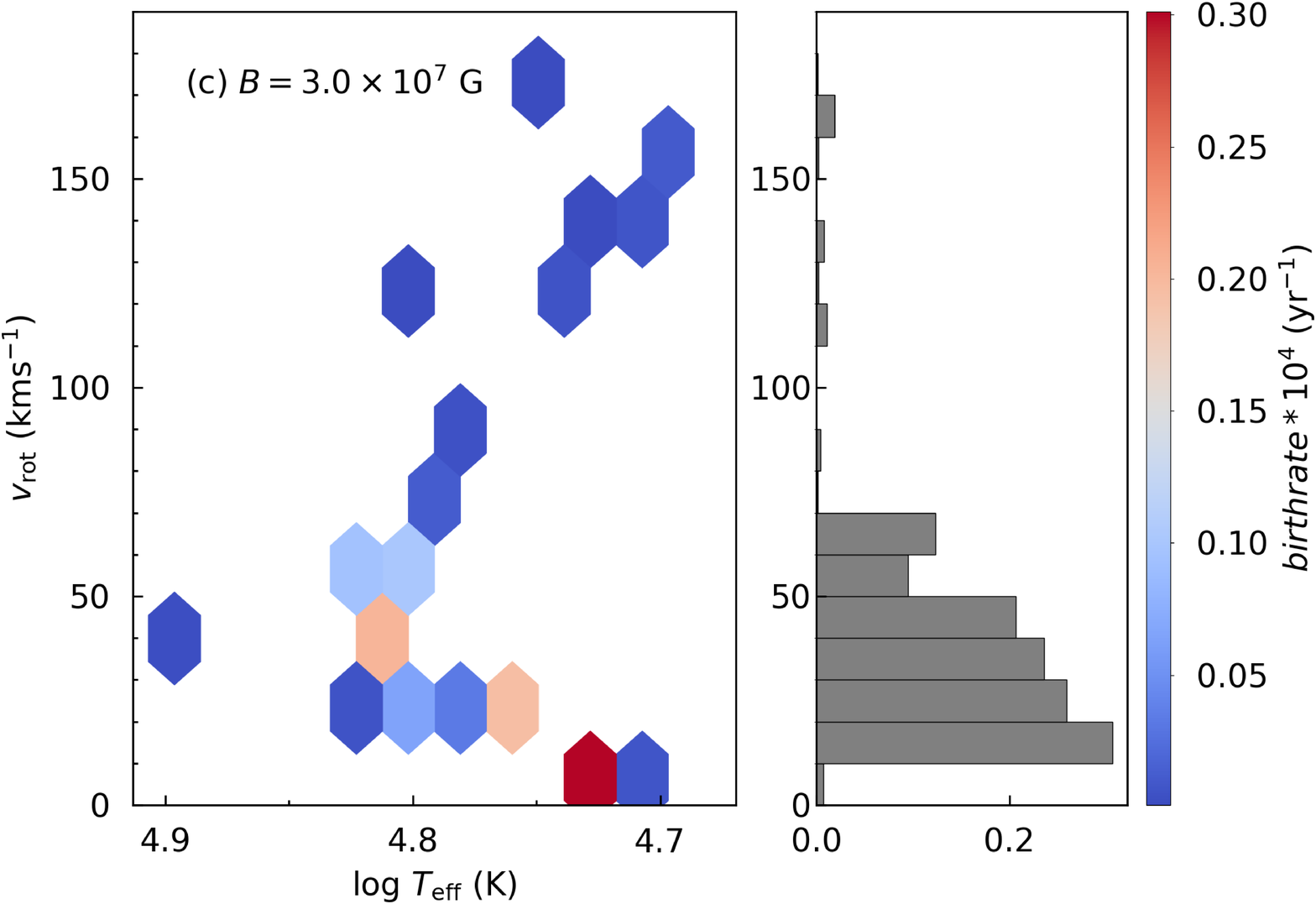}
	\caption{Same as Fig.~\ref{fig:f8} but with $\alpha = 1.0$. \label{fig:f9}}
\end{figure}

Our calculations also indicate that the surviving companions are characterized by relatively higher rotational velocities in the high-magnetic model than in the non- and intermediate-magnetic models. Figs.~\ref{fig:f8} and \ref{fig:f9} show the the rotational velocity $v_{\rm rot}$ vs. the effective temperature $T_{\rm eff}$ of the companion star with $\alpha = 0.5$ and 1.0, respectively. In the non- and intermediate-magnetic models generally $v_{\rm rot}\lesssim 150\, \rm kms^{-1}$ and a large fraction of the surviving companions have $v_{\rm rot} \lesssim 10\, \rm kms^{-1}$. In the high-magnetic model the surviving companions possess a rotational velocity $v_{\rm rot} \sim 10-170 \,\rm kms^{-1}$, which means that the binaries at the moment of SNe are relatively compact. This can be naturally explained by the initially narrow orbits of the progenitor binaries in the high-magnetic model.

\begin{figure}[htp]
	\centering
	\includegraphics[width=0.33\linewidth]{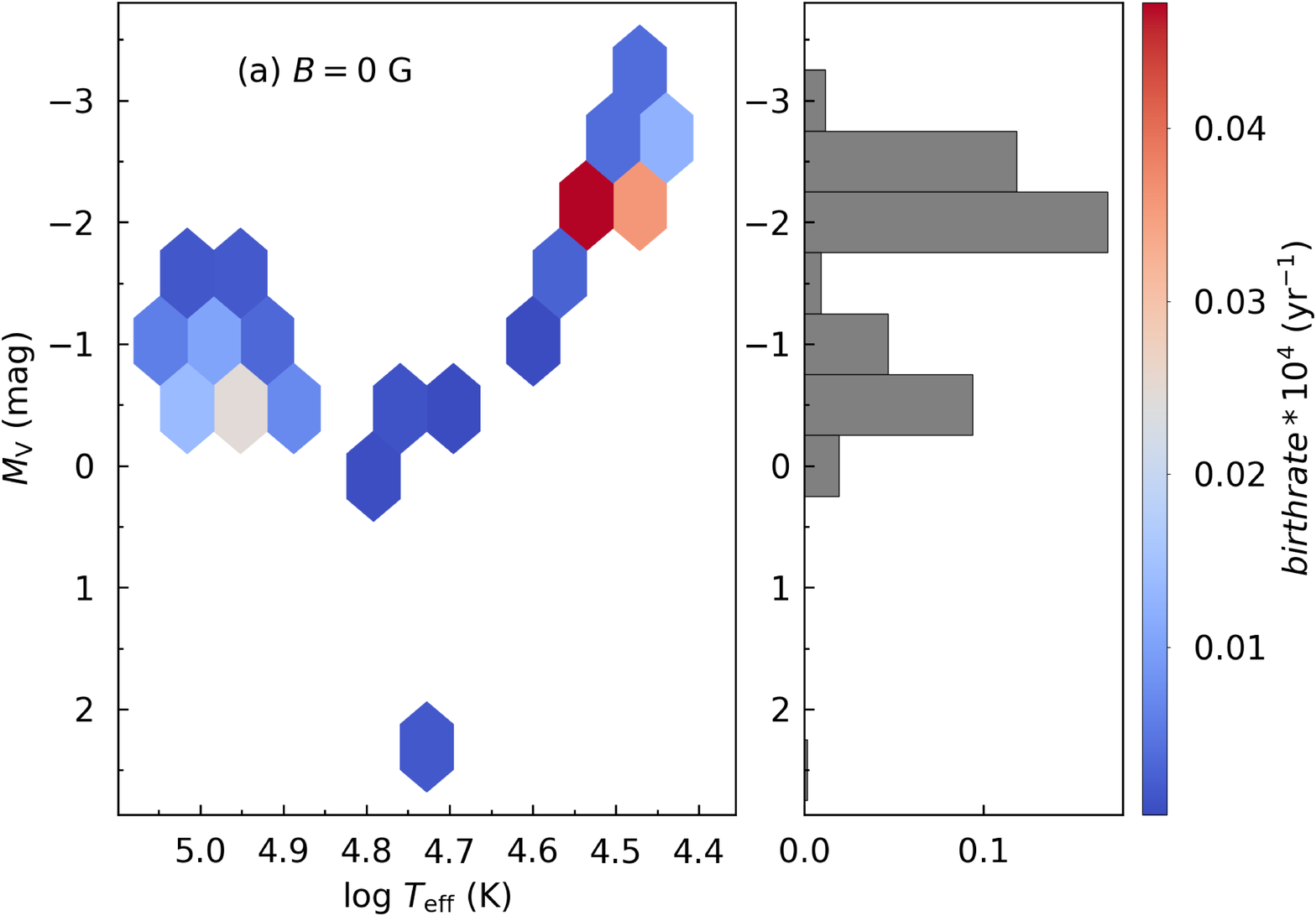}
	\includegraphics[width=0.33\linewidth]{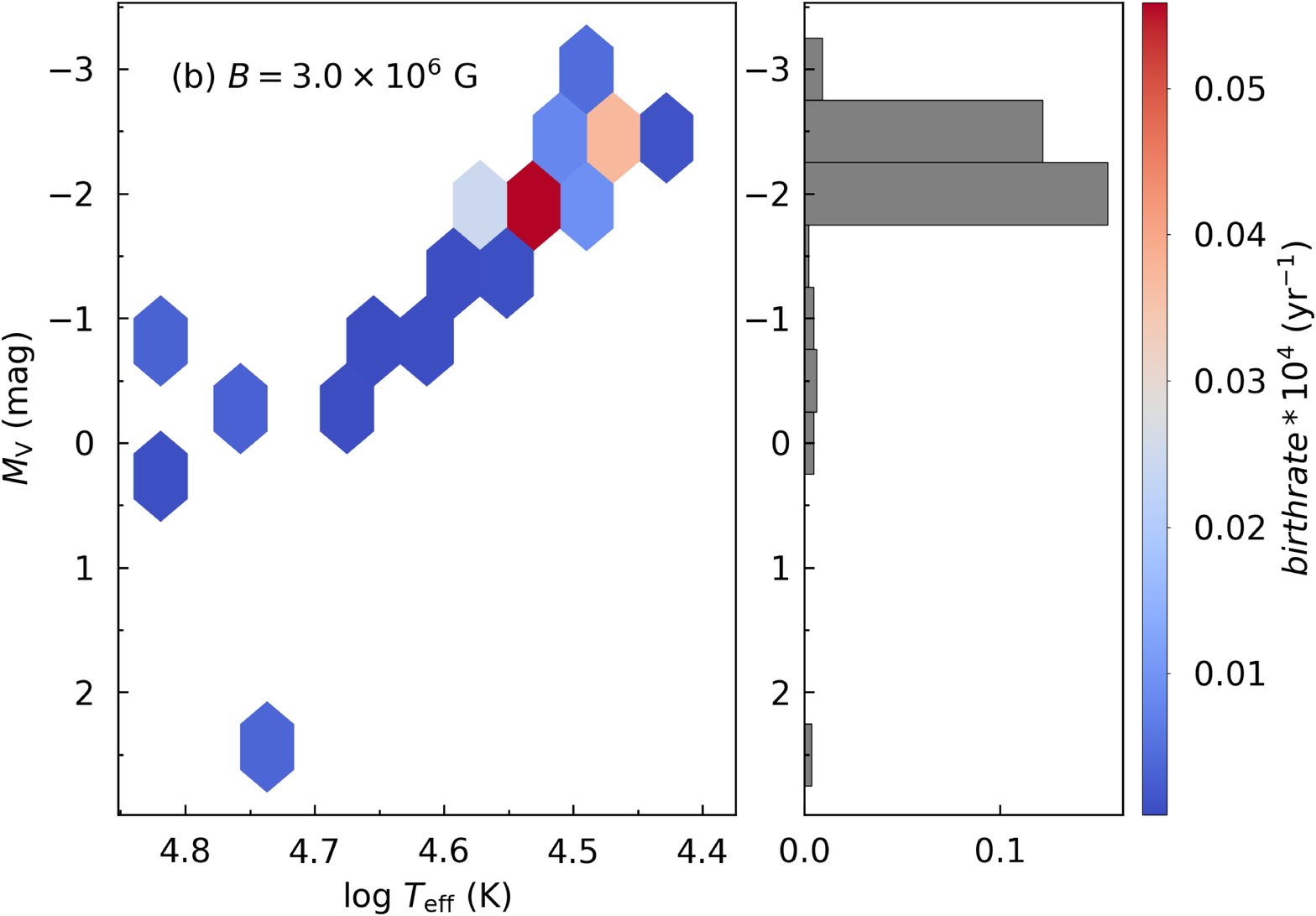}
	\includegraphics[width=0.33\linewidth]{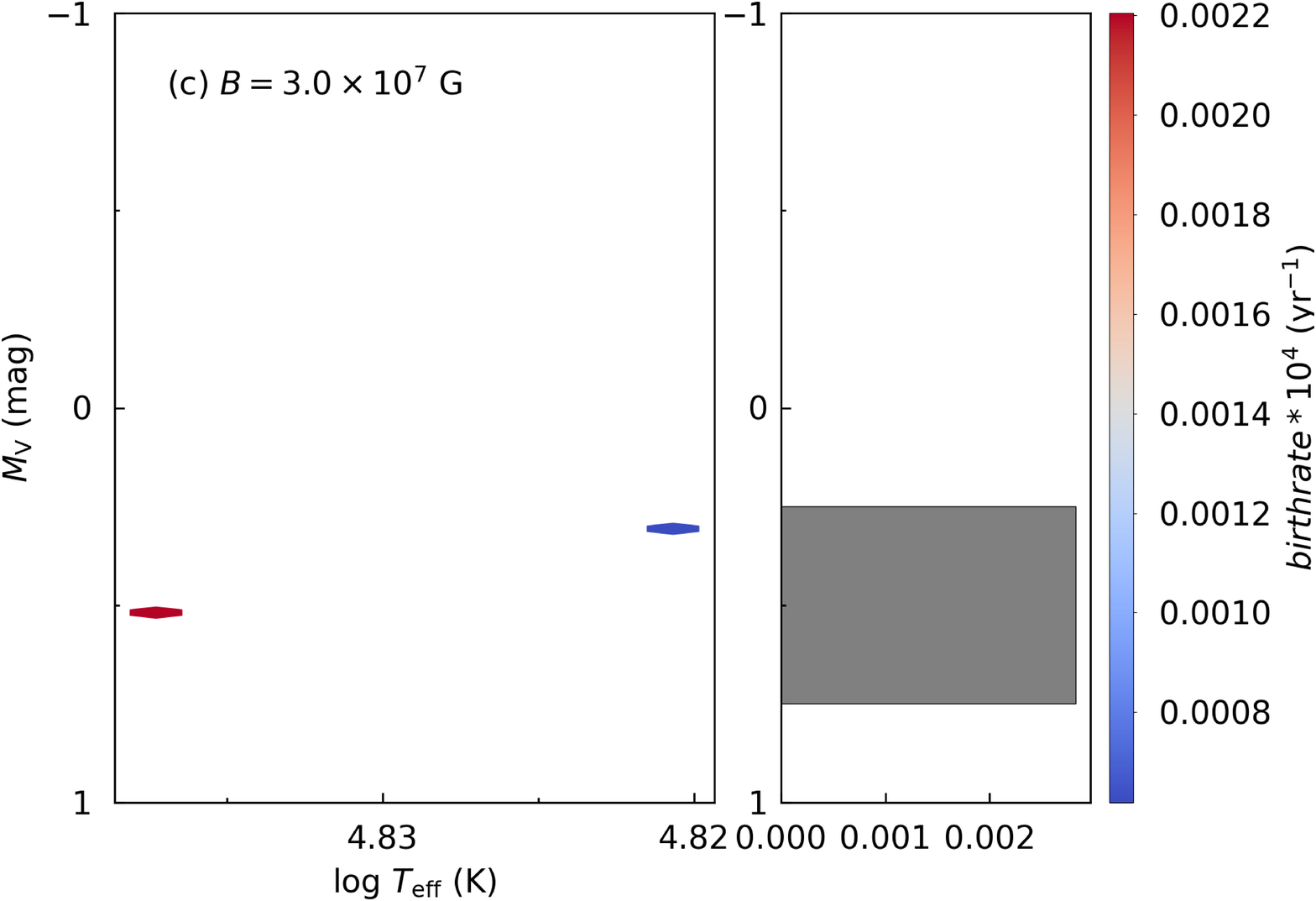}
	\caption{Same as Fig.~\ref{fig:f6} but with $\alpha = 0.5$ and the WL prescription for $\eta_{\rm He}$.\label{fig:f10}}
\end{figure}

\begin{figure}[htp]
	\centering
	\includegraphics[width=0.33\linewidth]{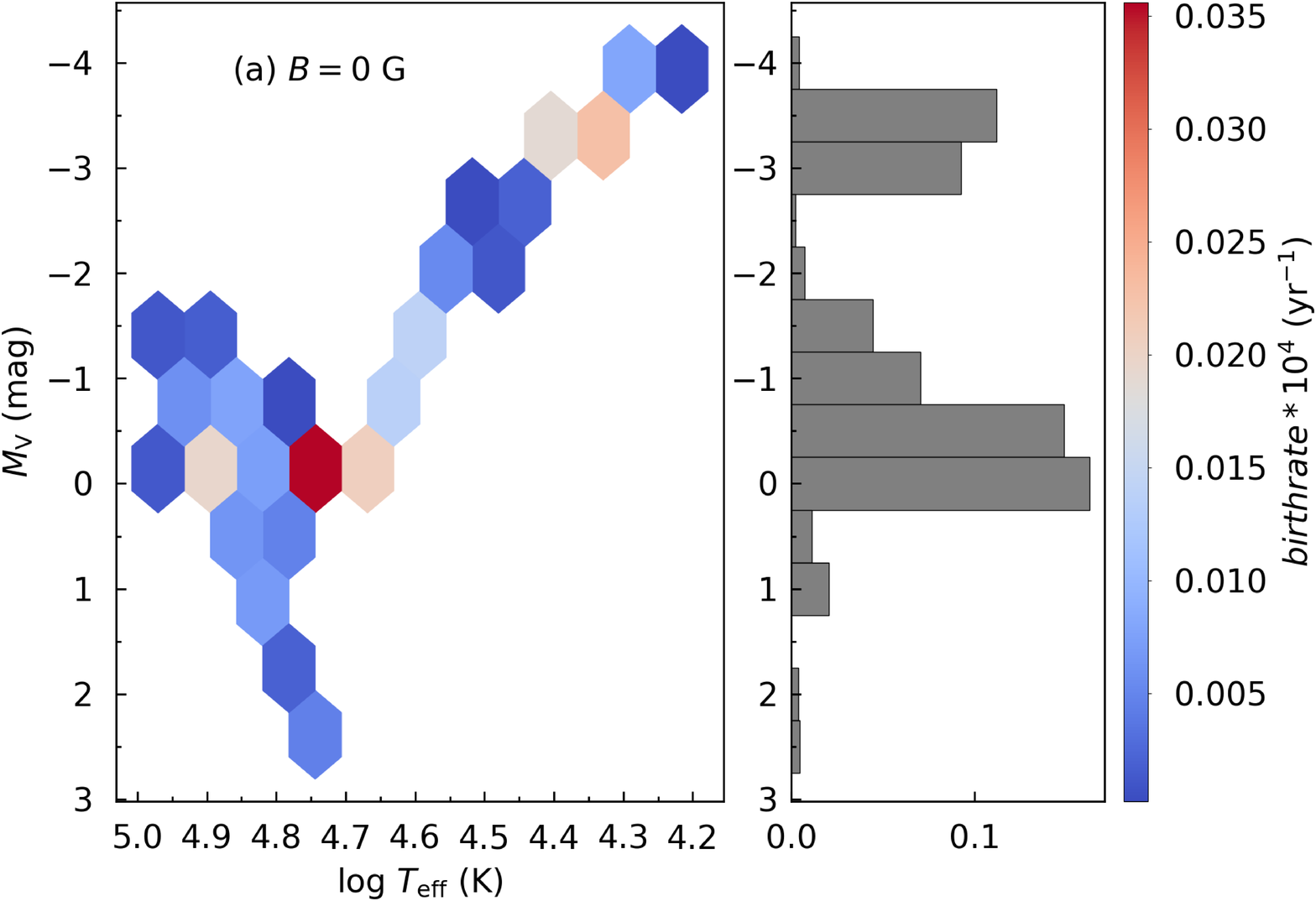}
	\includegraphics[width=0.33\linewidth]{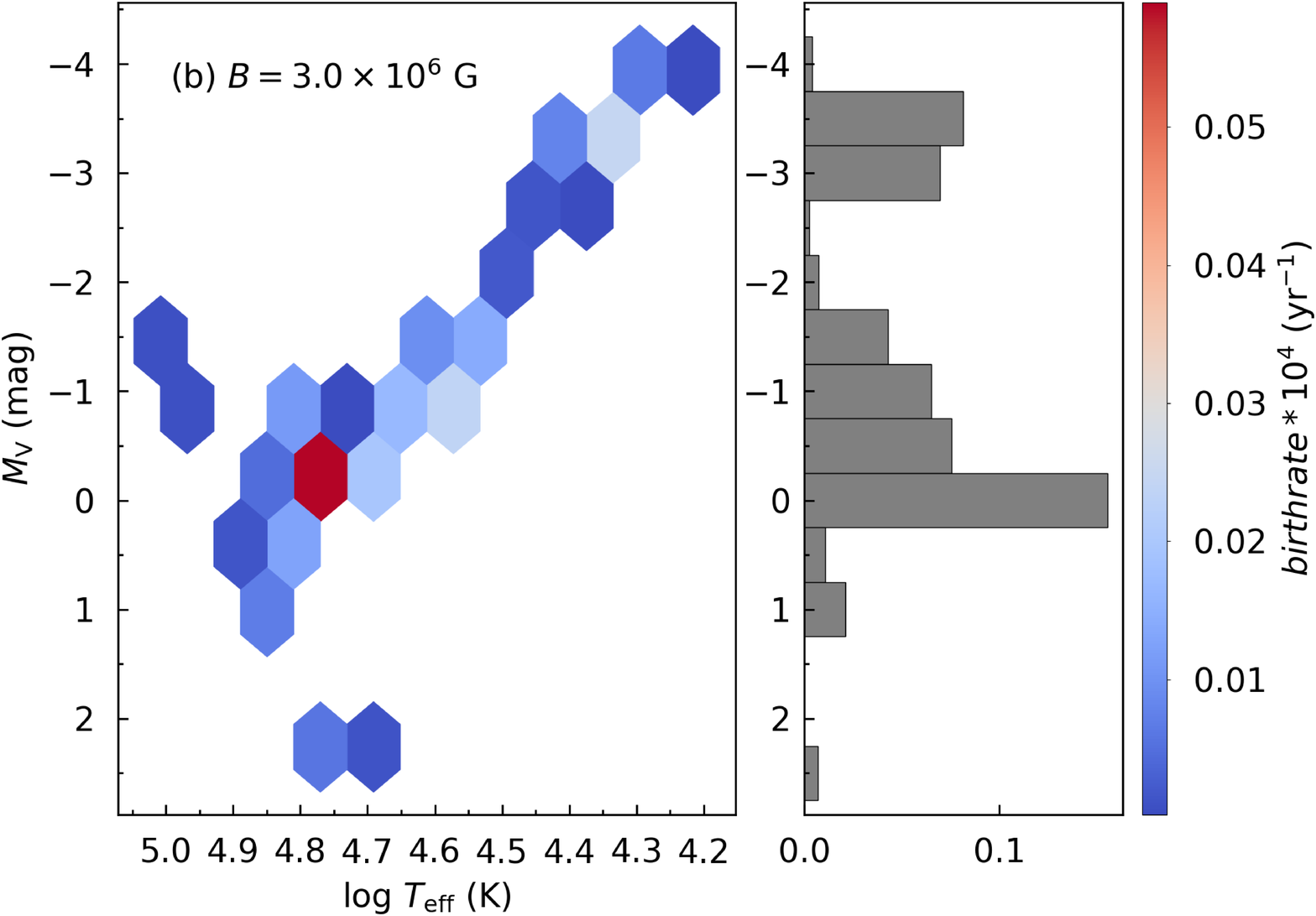}
	\includegraphics[width=0.33\linewidth]{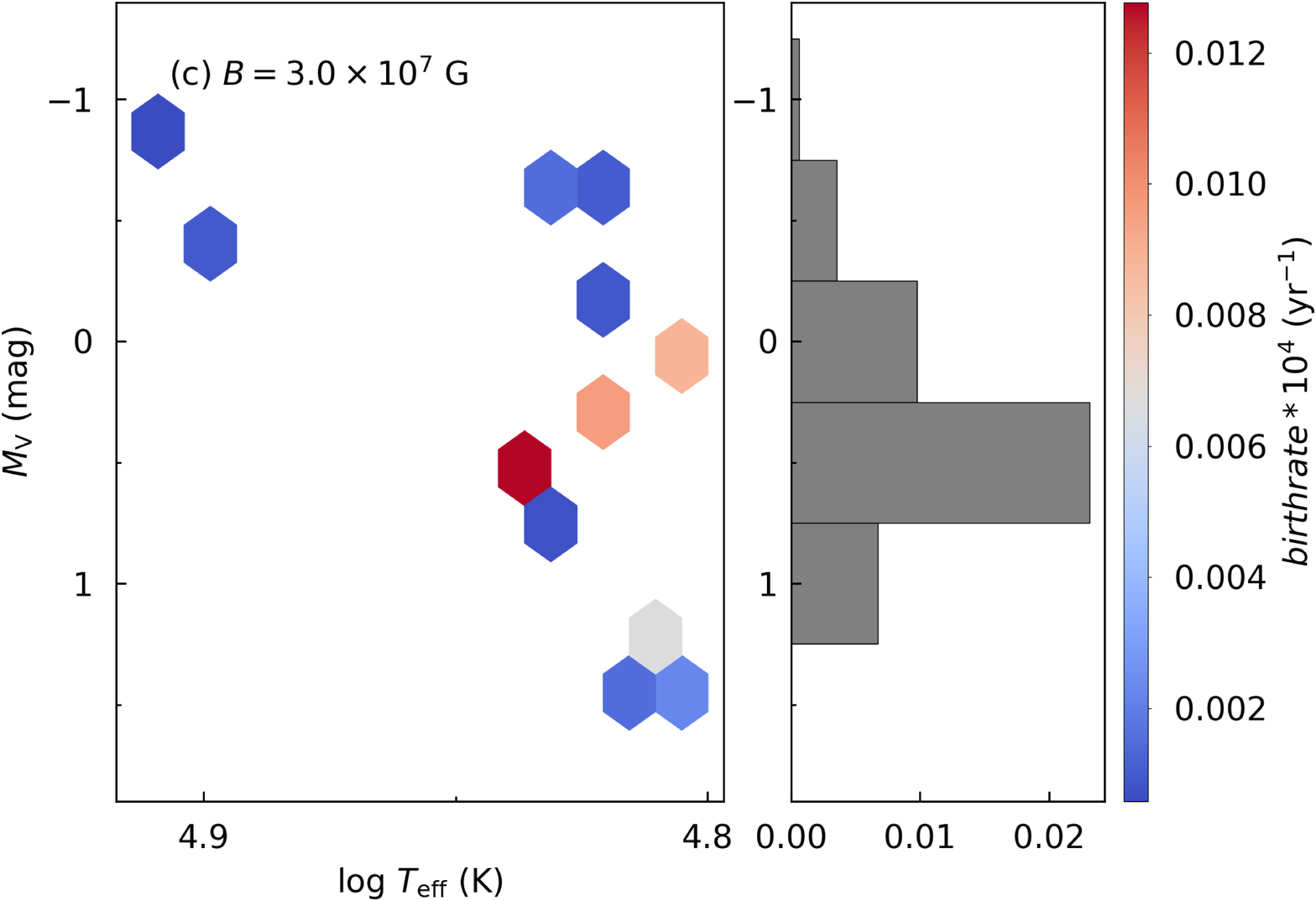}
	\caption{Same as Fig.~\ref{fig:f10} but with $\alpha = 1.0$.\label{fig:f11}}
\end{figure}

\begin{figure}[htp]
	\centering
	\includegraphics[width=0.33\linewidth]{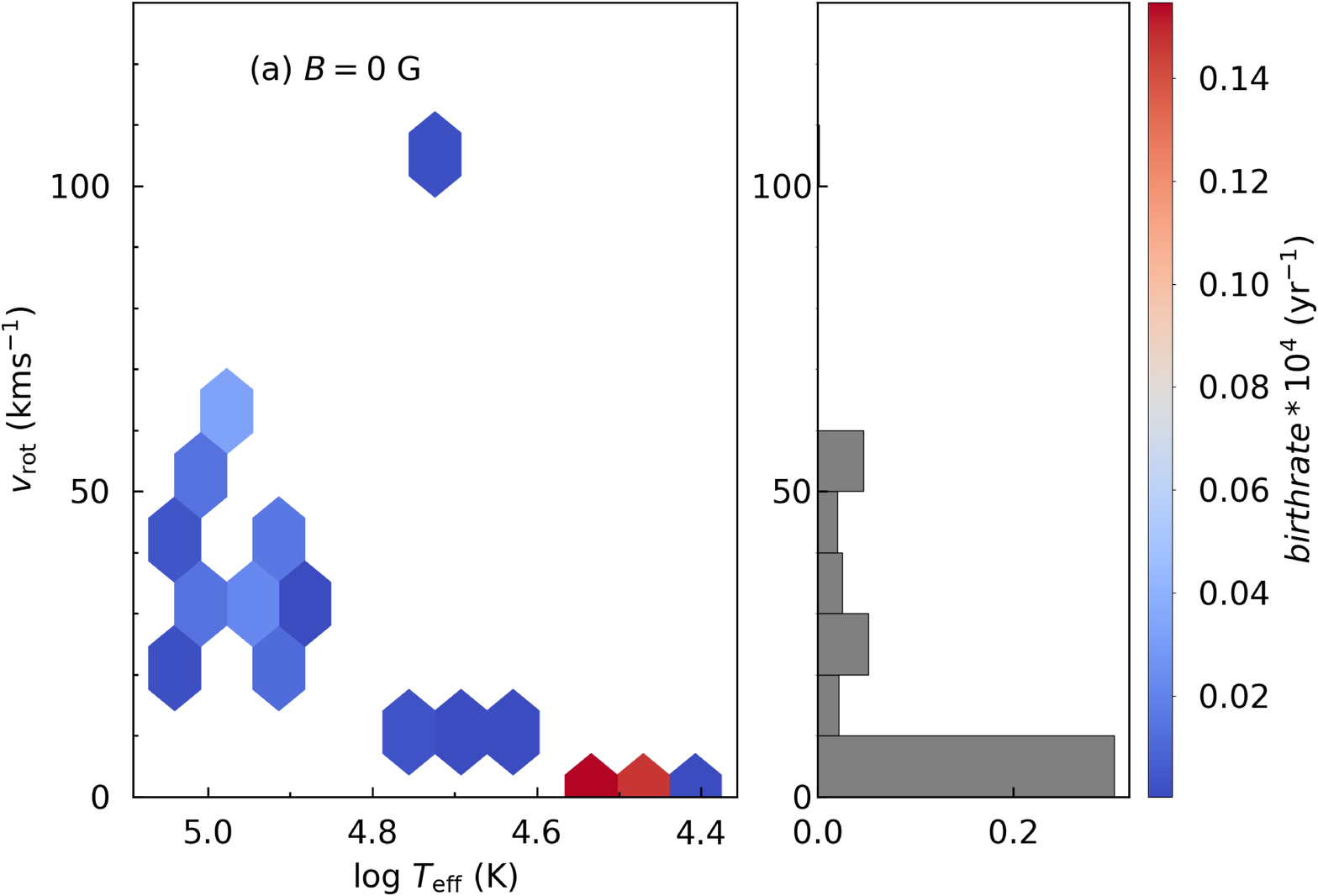}
	\includegraphics[width=0.33\linewidth]{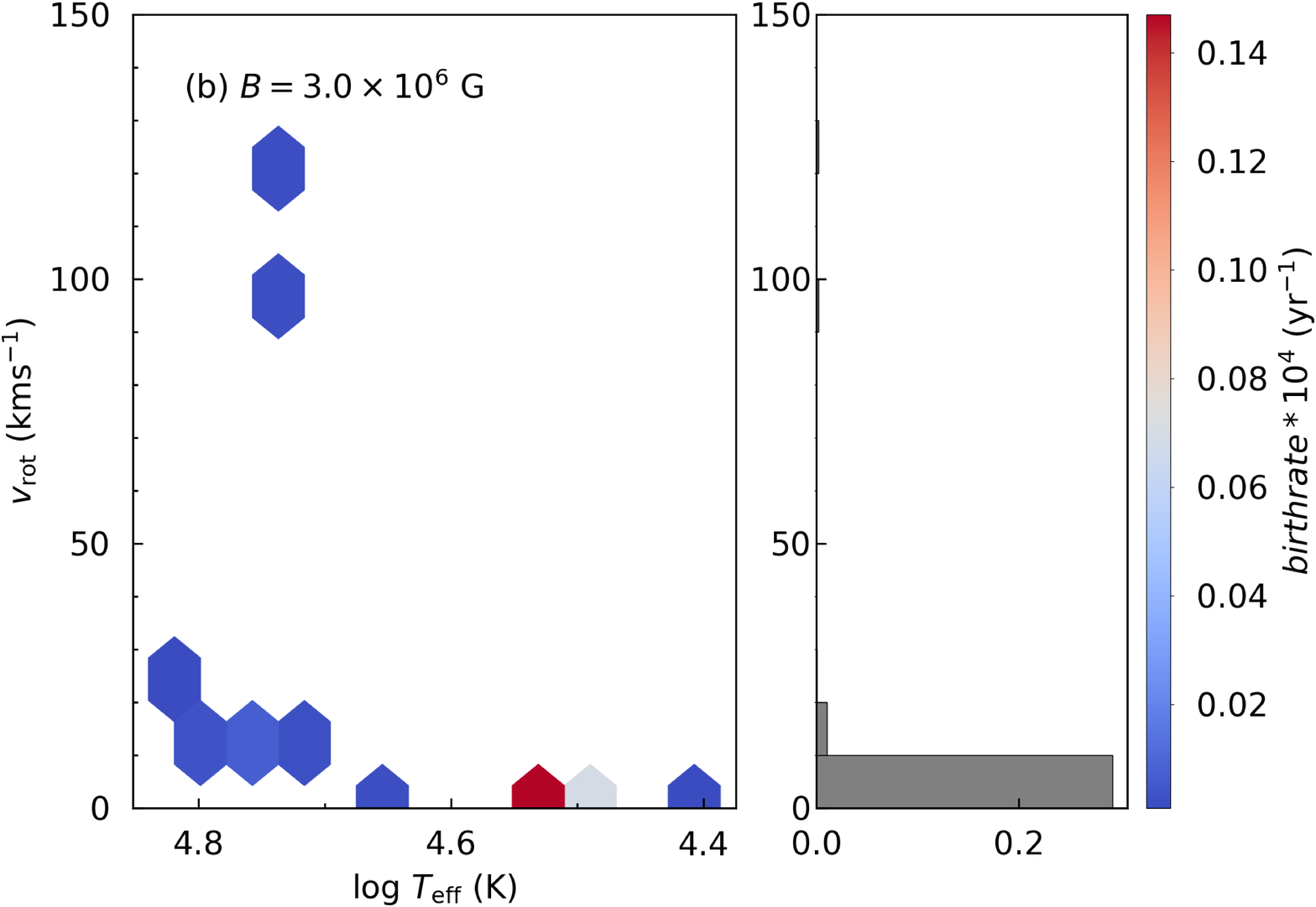}
	\includegraphics[width=0.33\linewidth]{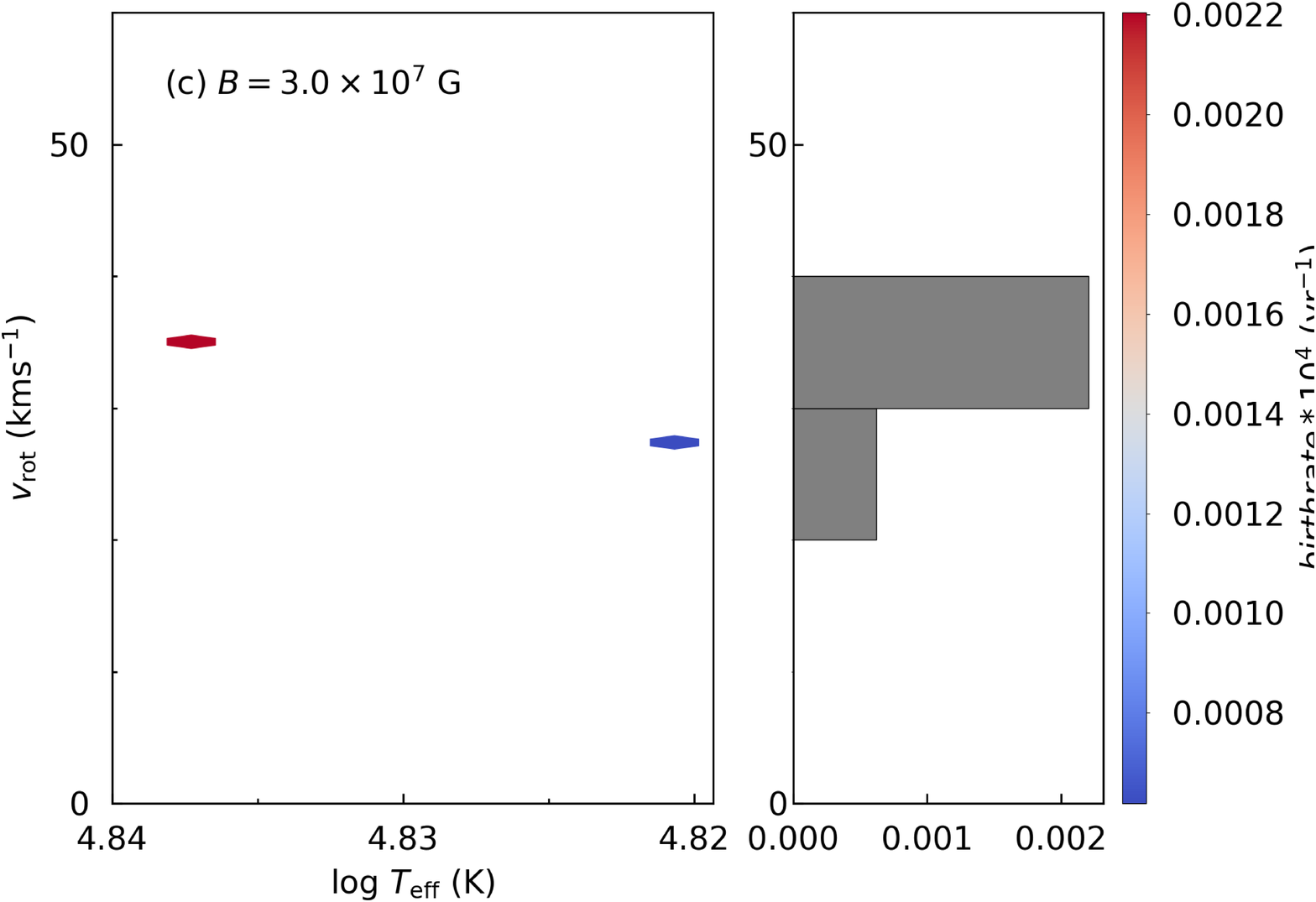}
	\caption{Same as Fig.~\ref{fig:f8} but with the WL prescription for $\eta_{\rm He}$.\label{fig:f12}}
\end{figure}

\begin{figure}[htp]
	\centering
	\includegraphics[width=0.33\linewidth]{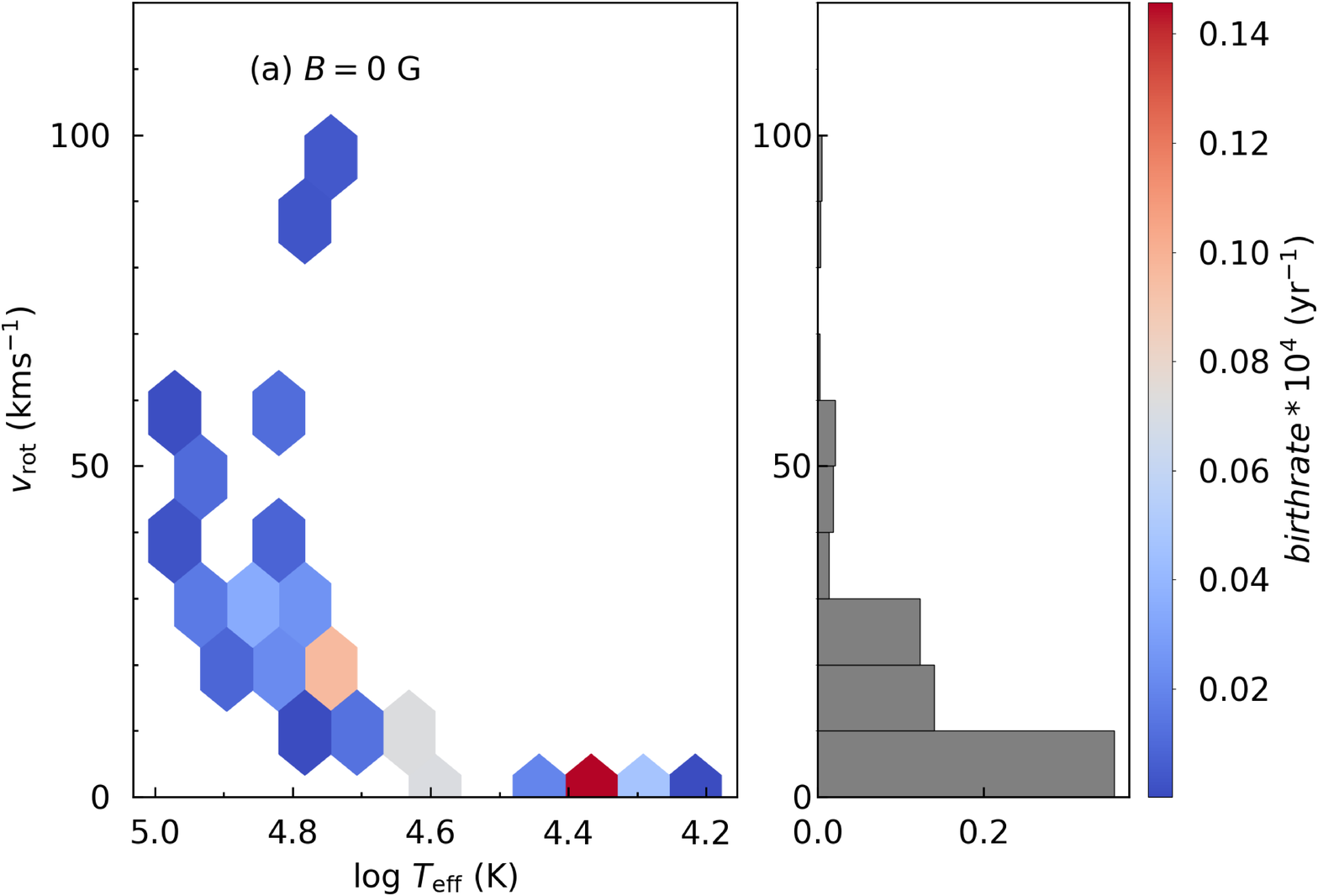}
	\includegraphics[width=0.33\linewidth]{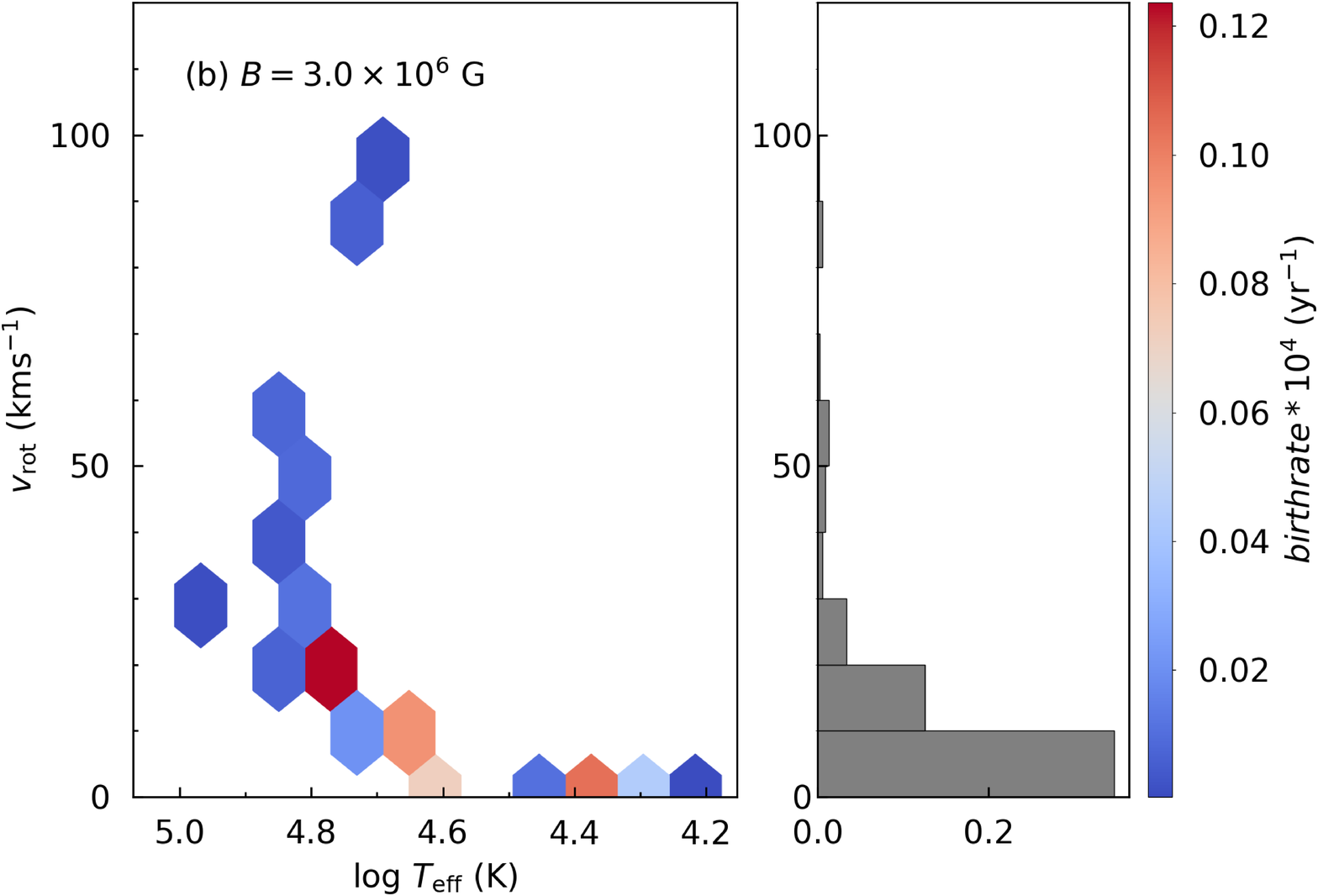}
	\includegraphics[width=0.33\linewidth]{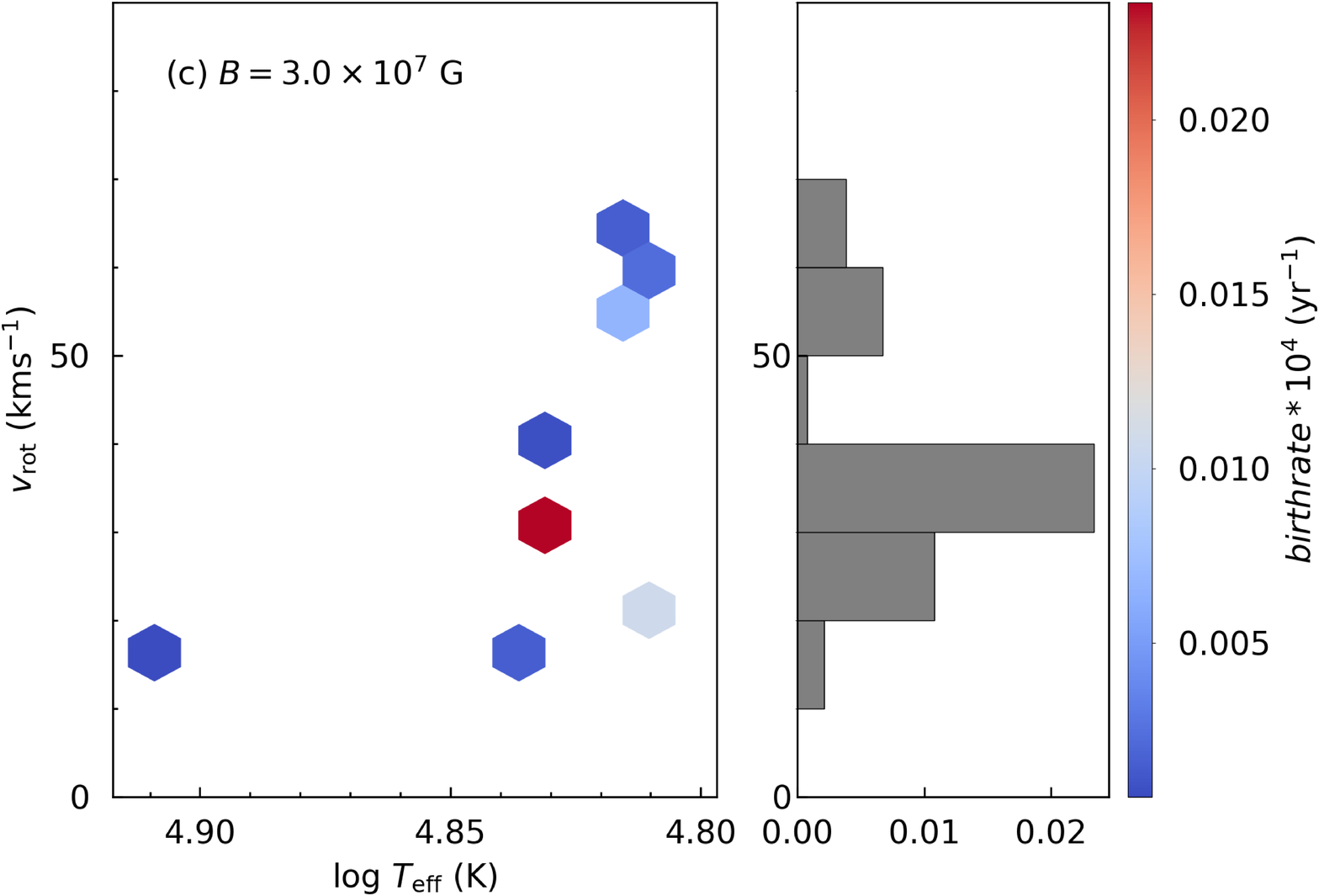}
	
	\caption{Same as Fig.~\ref{fig:f12} but with $\alpha = 1.0$.\label{fig:f13}}
\end{figure}

Figs.~\ref{fig:f10} - \ref{fig:f13} display the distributions of the surviving companions calculated with the WL prescription. Figs.~\ref{fig:f10} and \ref{fig:f11} show the distribution in the $M_{\rm V}-T_{\rm eff}$ plane, and Figs.~\ref{fig:f12} and \ref{fig:f13} in the $v_{\rm rot}-T_{\rm eff}$ plane. We take $\alpha=0.5$ in Figs.~\ref{fig:f10} and \ref{fig:f12}, and $\alpha=1.0$ in Figs.~\ref{fig:f11} and \ref{fig:f13}, respectively.
Generally speaking, Figs.~\ref{fig:f10} - \ref{fig:f13} follow the similar tendency as in Figs.~\ref{fig:f6} - \ref{fig:f9} with the KH prescription, but the birthrates are substantially lower than in the latter cases. In addition, the surviving companion stars are respectively slightly brighter and dimmer in the non-/intermediate-magnetic and high-magnetic models compared with those in Figs.~\ref{fig:f6} - \ref{fig:f9}.

\begin{figure}[htp]
	\centering
	\includegraphics[width=1.\linewidth]{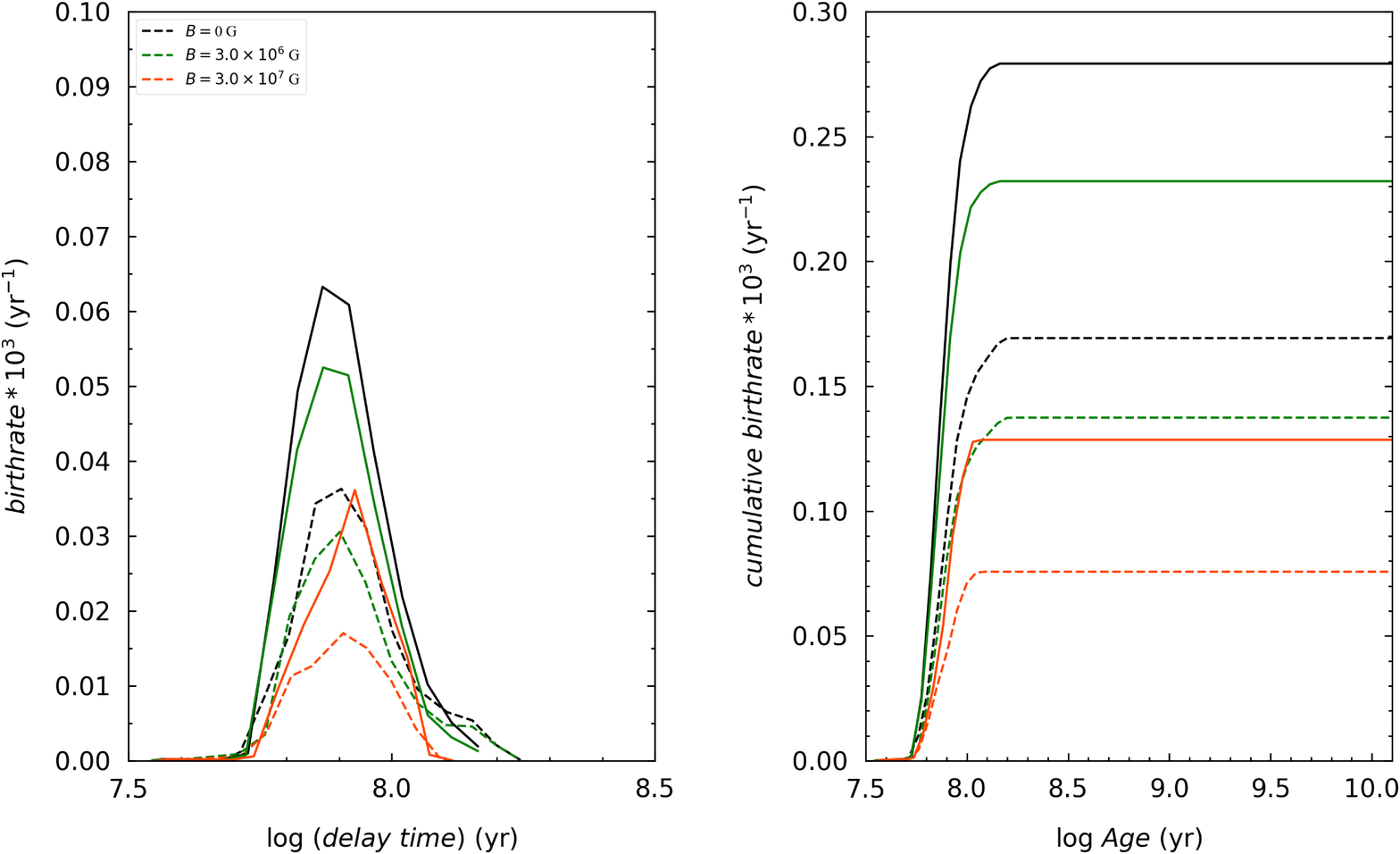}
	\caption{The left and right panels displays the DTD distributions and the cumulative birthrates of SNe Ia, respectively. Here we adopt the KH prescription for $\eta_{\rm He}$. The black, green, and orange-red curves represent the results of non-magnetic, intermediate-magnetic and high-magnetic models, respectively. The dashed and solid lines denote the results of $\alpha = 0.5$ and 1.0, respectively. \label{fig:f14}}
\end{figure}

\begin{figure}[ht]
	\centering
	\includegraphics[width=1.\linewidth]{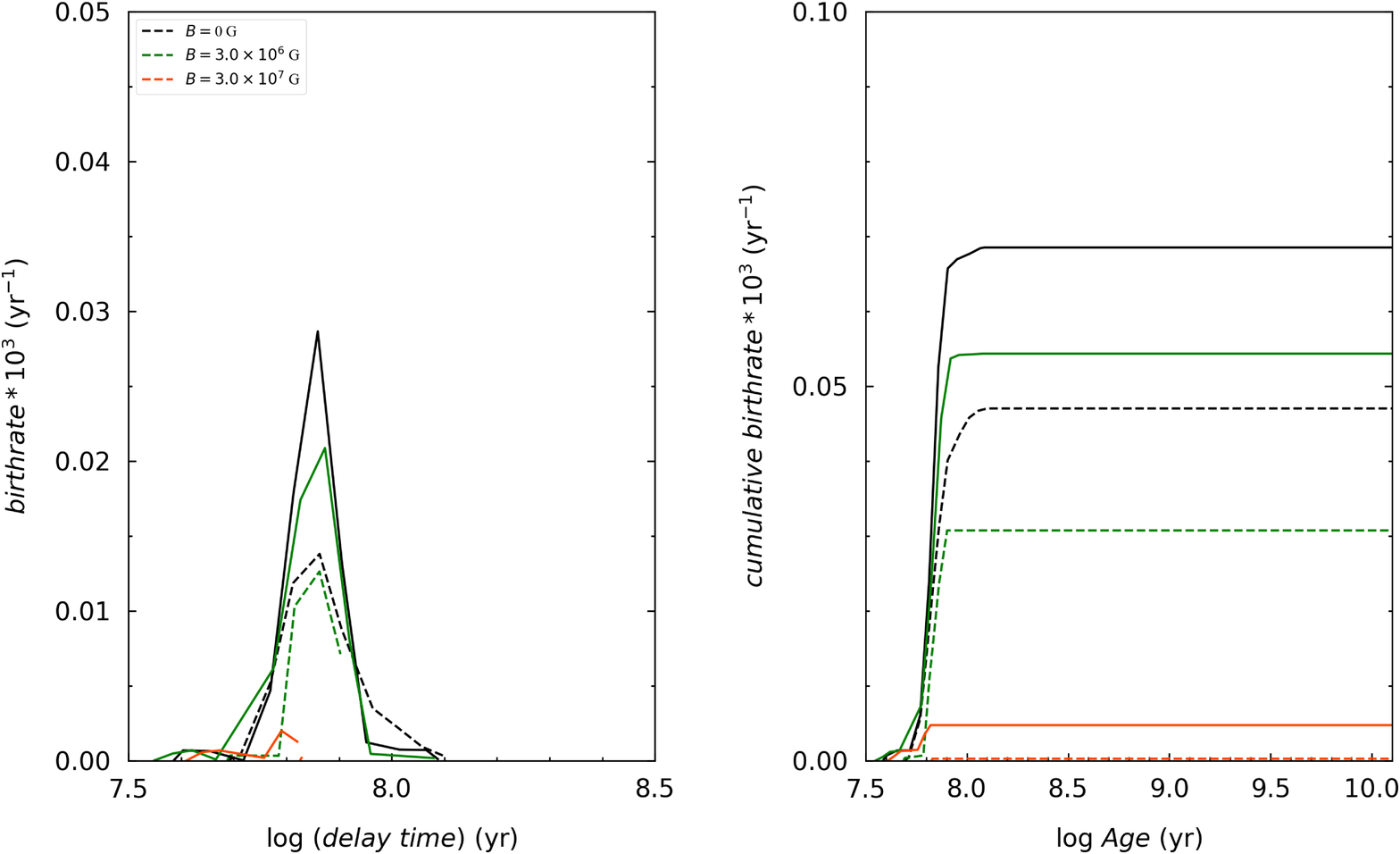}
	\caption{Same as Fig.~\ref{fig:f14} but with the WL prescription for $\eta_{\rm He}$. \label{fig:f15}}
\end{figure}

Figs.~\ref{fig:f14} and~\ref{fig:f15} show the DTD (left panel) and the cumulative birthrate evolution of SNe Ia (right panel) with the KH and WL prescriptions, respectively. The colors of the lines indicate different magnetic field strengths, and the results with $\alpha = 0.5$ and $1.0$ are displayed with the dashed and solid lines, respectively.
The DTD of SN Ia is defined as the time interval between the star formation to the SN explosion, which is dependent on the lifetimes of the progenitor binaries. The WD + He star channel links to relatively young populations. Figs.~\ref{fig:f14} and~\ref{fig:f15} show that the delay times are $\sim 35-160$ Myr (KH) and $\sim 35-126$ Myr (WL) in both the non- and intermediate-magnetic models, and $\sim 35-126$ Myr (KH) and $\sim 35-71$ Myr (WL) in the high-magnetic models, respectively. The relatively short delay times in the high-magnetic models is due to the lack of the progenitor systems with wider initial orbits compared with those in the non- and intermediate-magnetic models, as shown in Fig.~\ref{fig:f3}-\ref{fig:f5}.

In Fig.~\ref{fig:f14}, the predicted birthrates of SNe Ia are $\sim 0.17 \times 10^{-3}\, {\rm yr}^{-1}$ (with $\alpha = 0.5$) and $\sim 0.28 \times 10^{-3}\, {\rm yr}^{-1}$ (with $\alpha = 1.0$) in the non-magnetic models, $\sim 0.14 \times 10^{-3}\, {\rm yr}^{-1}$ (with $\alpha = 0.5$) and $\sim 0.23 \times 10^{-3}\, {\rm yr}^{-1}$ (with $\alpha = 1.0$) in the intermediate-magnetic model, and $\sim 0.08 \times 10^{-3}\, {\rm yr}^{-1}$ (with $\alpha = 0.5$) and $\sim 0.13 \times 10^{-3}\, {\rm yr}^{-1}$ (with $\alpha = 1.0$) in the high-magnetic models. 
The higher the $B$-fields, the smaller parameter space of the SN progenitors, and the lower the SN birthrates. In Fig.~\ref{fig:f15}, the birthrates with the WL prescription are significantly lower than in the corresponding models with the KH prescription, with the highest birthrate being $\sim 0.07 \times 10^{-3}\rm yr^{-1}$ in the non-magnetic model with $\alpha = 1.0$. We also see in Figs.~\ref{fig:f14} and \ref{fig:f15} that the birthrates predicted with $\alpha = 1.0$ are higher than those with $\alpha = 0.5$ because the primordial binaries are more likely to survive the CE evolution with higher $\alpha$.

The predicted Galactic SN Ia birthrate in the high-magnetic models is $\sim (0.08 - 0.13) \times 10^{-3}\,{\rm yr}^{-1}$ if all of the WDs are strongly magnetized, compared with $\sim (0.17-0.28) \times 10^{-3}\,{\rm yr}^{-1}$ in the non-magnetic models. Considering the fact that about $25\%$ of the WDs in binaries may be magnetic \citep{ferrario2015,pala2020}, the overall predicted birthrate is at most $\sim (0.15-0.24) \times 10^{-3} {\rm yr}^{-1}$.
This is significantly lower than the Galactic SN Ia birthrate \citep[$\sim 5 \times 10^{-3}\,{\rm yr}^{-1}$, e.g.,][]{patat2018}.
However, the magnetic WD + He star channel is still worth of investigation since the surviving companion stars predicted in this channel can be quite dim, providing a natural explanation for non-detection of bright surviving companions associated with some SNe Ia.

\section{Summary} \label{sec:summary}
In this paper, we investigate the contributions of the WD + He star binaries to SNe Ia by considering the WDs with non-, intermediate- and high-magnetic fields. With detailed binary evolution and BPS calculations we obtain the parameter space of potential SN Ia progenitors and the properties of the surviving companions. Compared with the case of non-magnetic accreting WDs, magnetically confined mass accretion makes it less possible for the WDs to steadily accumulate mass owing to strong wind mass loss, meaning that the influence of the magnetic confinement on He-accreting WDs is quite different from that on H-accreting WDs. This is because the helium-ignition pressures $P_{\rm b}$ is typically $\sim 10 -1000$ times higher than that for hydrogen-ignition, and increases with decreasing mass transfer rates. Thus it is difficult for the magnetic confinement to take effect and enhance the He-rich mass accumulation efficiency $\eta_{\rm He}$ in systems with relatively low $\dot{M}$. Our calculations show that  the magnetic confinement does lead to different characteristics of the surviving companion stars, such as the mass, luminosity and rotational velocity distributions. However, the predicted overall birthrates are still significantly lower than the Galactic SN Ia birthrate. This further strengthens that WD + He binaries, even with the effect of magnetic fields of the WDs taken into account, are not likely to be the primary progenitors of SNe Ia.

\normalem
\begin{acknowledgements}
We are grateful to an anonymous referee for careful reading the manuscript and helpful comments. This work was supported by the National Key Research and Development Program of China (2016YFA0400803), the Natural Science Foundation of China under grant No. 11773015, 12041301 and Project U1838201 supported by NSFC and CAS.
\end{acknowledgements}
  
\bibliographystyle{raa}
\bibliography{ms2021-0313}

\end{document}